\documentclass[lettersize,journal]{IEEEtran}
\usepackage{amsmath,amsfonts}
\usepackage{algorithmic}
\usepackage{array}
\usepackage[caption=false,font=normalsize,labelfont=sf,textfont=sf]{subfig}
\usepackage{textcomp}
\usepackage{stfloats}
\usepackage{url}
\usepackage{verbatim}
\usepackage{graphicx}
\usepackage{booktabs}
\usepackage{multirow}
\usepackage{tikz}
\usetikzlibrary{arrows.meta,positioning,fit,calc}
\def\BibTeX{{\rm B\kern-.05em{\sc i\kern-.025em b}\kern-.08em
    T\kern-.1667em\lower.7ex\hbox{E}\kern-.125emX}}
\usepackage{balance}
\usepackage{cite}
\usepackage{subfig}
\usepackage{xcolor}
\usepackage{tikz}
\usetikzlibrary{
	arrows.meta,    % 箭头样式（Latex 箭头）
	positioning,    % 位置定位（below=6mm of feat 等）
	fit,            % 包裹多个节点（fit=(feat)(dec)）
	calc            % 坐标计算（$(igfa.west)!0.5!(igna.west)$）
}

\usepackage[normalem]{ulem} % 提供 \sout
\NewDocumentCommand{\added}{O{}m}{\textcolor{blue}{#2}}
\NewDocumentCommand{\deleted}{O{}m}{\textcolor{red}{\sout{#2}}}
\NewDocumentCommand{\replaced}{O{}mm}{\textcolor{blue}{#2}\textcolor{red}{\sout{#3}}}
\begin{document}
\title{IGDA-ICMH: an {Importance Guided} Feature and Network {Dual Adaptation} Framework towards {Image Compression for Machine and Human} Vision}
%\title{IGDA-ICMH: an \textit{Importance Guided} Feature and Network \textit{Dual Adaptation} Framework towards \textit{Image Compression for Machine and Human} Vision}
\author{Yanbo Gao, Xinyu Zhao, Shuai Li\textsuperscript{*},~\IEEEmembership{Senior Member,~IEEE}, Xun Cai, Mao Ye,~\IEEEmembership{Senior Member,~IEEE}, \\Li Li,~\IEEEmembership{Senior Member,~IEEE}
	
\thanks{Manuscript received Apr. 2, 2026.}
\thanks{This work was supported in part by the National Natural Science Foundation of China (No. 62572275, U24B20132 and 62271290), and the Shandong Provincial Natural Science Foundation (No. ZR2024LZN021, ZR2025QB44 and ZR2025MS1074. }%        
%\thanks{\textsuperscript{*}Corresponding author: Shuai Li. e-mail: shuaili@sdu.edu.cn}%
%\thanks{Zhiyuan Yang, Shuai Li are with the School of Control Science and Engineering, Shandong University, Jinan 250100, China, and also with the Key Laboratory of Machine Intelligence and System Control, Ministry of Education, Jinan 250100, China.}%
%\thanks{Yanbo Gao is with the School of Software, Shandong University, Jinan 250100, China, and also with Shandong University-WeiHai Research Institute of Industrial Technology, Weihai 264209, China.}%
%\thanks{Chuankun Li is with the State Key Laboratory of Dynamic Testing Technology and School of Information and Communication Engineering, North University of China, Taiyuan 030051, China.}%
\thanks{Y. Gao, X. Zhao and X. Cai are with the School of Software, Shandong University, Jinan 250100, China, and also with Shandong University-WeiHai Research Institute of Industrial Technology, Weihai 264209, China. E-mail:  ybgao@sdu.edu.cn.
\par S. Li is with the School of Control Science and Engineering, Shandong University, Jinan 250100, China. E-mail:  shuaili@sdu.edu.cn.
%\par Jiande Sun is with the School of Computer Science and Artificial Intelligence, Shandong Normal University, Jinan 250358, China.
\par  M. Ye is with University of Electronic Science and Technology of China, Sichuan, China. E-mail: cvlab.uestc@gmail.com.%
\par  L. Li is with University of Science and Technology of China (USTC), Hefei, Anhui, China. E-mail: lil1@ustc.edu.cn.}%
}

\maketitle

\begin{abstract}
With the surge of machine vision applications, image compression  for both machine and human vision (ICMH) has become an important research area. Existing studies mainly focus on task-specific adaptation or multi-task joint optimization, yet they often overlook the unequal contributions of different features and network layers to downstream machine vision tasks. To address this issue, we first quantitatively investigate the impact of different feature and network layers on machine vision performance, providing a principled understanding of their importance. Based on this analysis, an \textit{Importance Guided} feature and network \textit{Dual Adaptation} (IGDA) framework is developed for ICMH. It adopts implicit neural representation as the basic architecture and designs an efficient adaptation structure with both feature and network weight adaptation. An Importance-Guided Network Adaptation (IGNA) module is developed with a low-rank adaptation strategy to selectively adjust key network layers according to their importance in different tasks. An Importance-Guided Feature Adaptation (IGFA) module is developed with a low-rank and Kronecker product-based transform to explore the non-local property of the features in order to reduce the number of adaptation parameters. Extensive experiments on representative machine vision tasks, including image classification and object detection, demonstrate that our IGDA-ICMH framework achieves significantly better performance than the existing implicit neural representation-based  methods for ICMH.
\end{abstract}

\begin{IEEEkeywords}
Learned Image Compression, Image Coding for Machine Vision, Implicit Neural Representation.
\end{IEEEkeywords}

\section{Introduction}
\IEEEPARstart{T}{he} rapid development of computer vision technology has raised higher requirements for the machine vision tasks such as image classification, object detection, and instance segmentation. It also brings a significant increase in the demand for visual data storage and transmission. The traditional vision processing pipeline, directly storing raw data and using it for machine vision tasks,  is becoming inefficient. On the other hand, to reduce the storage of image data, image compression is widely used. The existing image compression methods, including the traditional image compression methods and deep learning-based methods, are optimized for human vision. These compression methods may lead to the loss of semantic information contained in an image and dramatically reduce the accuracy of downstream vision tasks. To meet the demands of both machine and human vision tasks, image compression for machine and human vision (ICMH), with a unified compression codec, is highly desired and attracts increasing interest.

The existing ICMH methods can be generally classified into three categories. The first approach aims at learning a unified feature representation and uses different decoders for different tasks. In \cite{feng2023semantically,bai2022towards}, a shared encoder is trained to extract common representations, while task‑specific decoders are developed to support human and machine vision tasks. The second approach  usually adopts a scalable codec, using a base layer carrying semantic information for machine vision and the enhancement layer carrying additional visual details for human vision. In \cite{choi2022scalable,liu2021semantics}, a hierarchical latent representation model is adopted in which the latent space is partitioned into two layers, separately encoding machine-oriented and human-oriented information. Since the scalable codec is trained using the pre-defined machine vision tasks, it lacks flexibility to adapt to new machine vision tasks. The third approach employs a lightweight trainable module to adapt a pre-trained image codec optimized for human vision to various machine vision tasks. In \cite{feng2023prompt}, Prompt-ICM employs task-driven prompts to guide an image codec for multiple tasks. In \cite{li2024image}, feature adaptation is realized by inserting a lightweight adapter that modulates features in both spatial and frequency domains, effectively filtering task‑irrelevant information.

Although the above ICMH methods have improved the performance over training a completely different image compression codec for different tasks, there is a lack of investigation on the effects of different layers or features of the network on the different tasks. In other words, there is no systematic understanding of which network layers and features in the compression model for human vision play a decisive role in the downstream machine vision task performance. This presents a bottleneck for developing more efficient and lightweight improvements to the model. 

On the other hand, Implicit Neural Representations (INRs) have emerged as a powerful alternative for representing signals. Unlike traditional codecs, INRs parameterize a signal as the weights of a small neural network. In \cite{ladune2023cool}, Cool-Chic uses a multi-resolution feature grid and a small network to represent an image. Then the feature grid and the network are quantized and entropy coded into a bitstream for storage and transmission. 
%修改修改
%Despite its success in the compression performance, existing INR methods are mostly developed for human vision-oriented compression, and have seldomly been explored for machine vision yet.
%修改修改
Despite its success in compression performance, existing INR methods are primarily designed for human vision-oriented compression, and its application in the context of ICMH remains very limited. In \cite{li2024human}, INR is preliminarily adopted for ICMH, explored its applicability to ICMH. In view of this, our method adopts an INR-based paradigm, and further explore an effective INR-based ICMH method. In this paper, we first design a systematic importance analysis experiment to quantitatively evaluate the contribution of each layer and feature to the machine vision tasks based on INR. 

\begin{figure}[!t]
	\centering
	
	\subfloat[Results of different layers.]{
		\includegraphics[width=0.8\columnwidth]{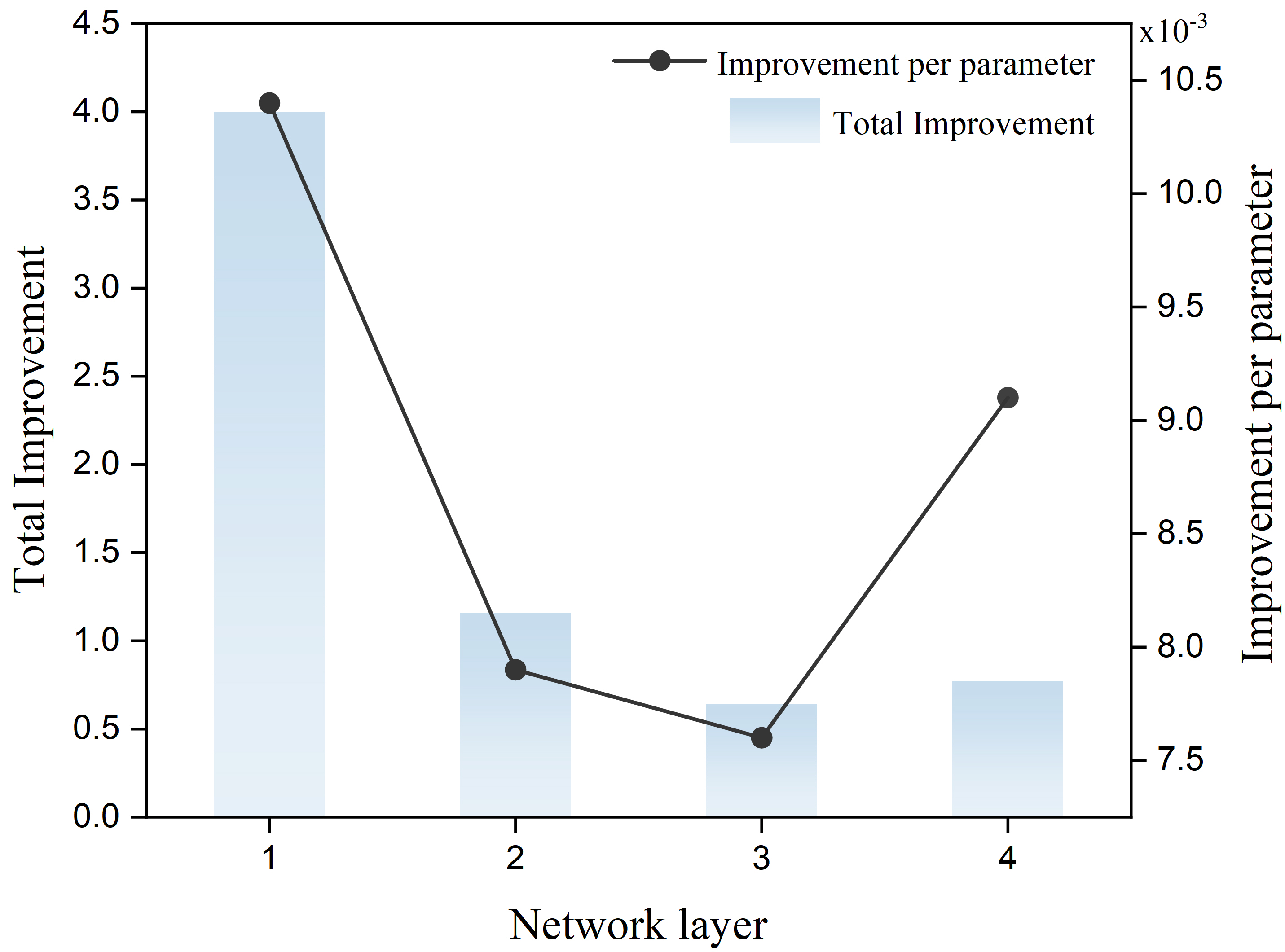}
		\label{fig:2a}
	}
	
	\par\medskip
	
	\subfloat[Results of features at different resolutions.]{
		\includegraphics[width=0.8\columnwidth]{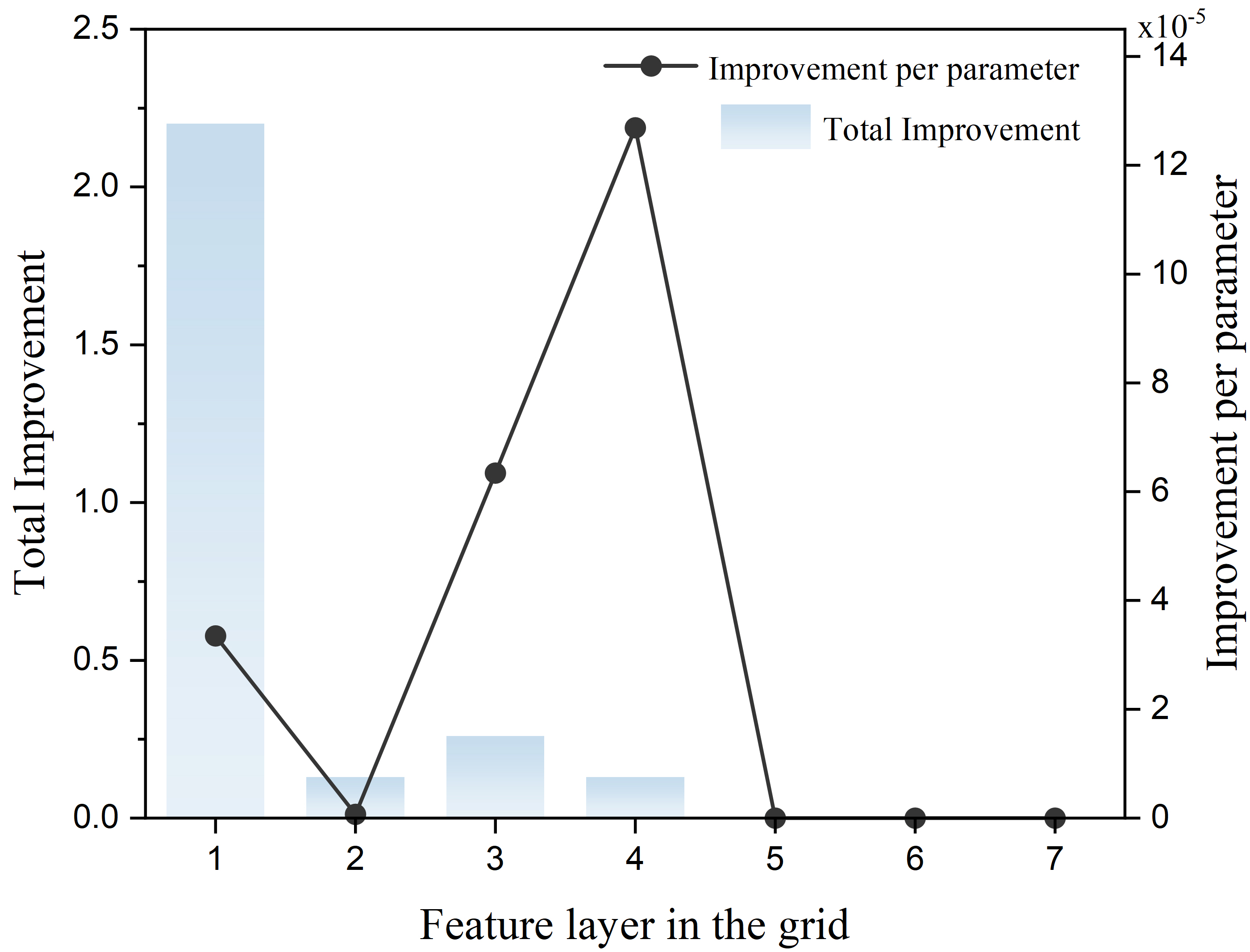}
		\label{fig:2b}
	}
	
	\caption{Illustration of total accuracy improvement per layer and average accuracy improvement per parameter, indicating different importance of the layers and features to the final performance.}
	\label{fig:DifferentModules}
\end{figure}

Specifically, two importance analysis experiments are conducted, where the first one evaluates the effect of different layers and the second one evaluates the effect of features at different resolutions in the feature grid. The pre-trained implicit neural framework Cool-Chic for human vision is adopted as the base codec and the image classification task is used as the downstream machine vision task. 
%修改修改
%Each feature and layer of the base codec is sequentially fine-tuned with others frozen, to assess how each feature and layer affect the classification accuracy. Fig. \ref{fig:DifferentModules} shows the average accuracy improvement per parameter for each feature and layer adaptation. It can be observed that although the classification accuracy has improved with finetuning a feature or layer, the results are significantly different for different features and layers.
%修改修改
Firstly, we sequentially unfreeze and fine-tune one layer of the base codec at a time, while keeping all other parameters fixed, to assess how each layer affects the classification accuracy. Fig. \ref{fig:DifferentModules}\subref{fig:2a} shows the average accuracy improvement per parameter for each layer adaptation. It can be observed that although the classification accuracy has improved with finetuning a layer, the results are different for different layers. Specifically, the first and fourth layers achieve better tradeoff than the second and third layers when considering the average improvement in classification accuracy per parameter. {It is worth noting that the bar (total accuracy improvement) and the curve (average accuracy improvement per parameter) shown in Fig. \ref{fig:DifferentModules}\subref{fig:2a} seem inconsistent due to the different number of parameters in different layers.}

Secondly, the features of different resolutions in the multi-resolution feature grid are also sequentially tested. The results are shown in Fig. \ref{fig:DifferentModules}\subref{fig:2b}, and similar to the results of layer adaptation, the different resolutions of features also demonstrate different effects on the final classification accuracy. Specifically, fine-tuning the fourth feature layer shows the greatest improvement in average classification accuracy per parameter. On the other hand, fine-tuning the sixth feature layer and the seventh feature layer do not improve the classification accuracy, leading to their indifference to the human vision and machine vision tasks. This makes sense since the low-resolution feature usually corresponds to the high-level semantic information useful for both tasks. Therefore, in adapting the network for different tasks, carefully selecting the appropriate layer and features for adaptation, in order to achieve the best tradeoff between the number of trainable parameters and the performance, is required.

Motivated by the above analysis, we propose an importance guided feature and network dual adaptation framework for ICMH. It first identifies the bottlenecks in improving pretrained human vision codecs for machine vision. Then, efficient adaptation strategies based on low-rank decomposition and Kronecker product are developed. Moreover, the parameter budget is adaptively assigned to different feature and network layers according to their importance to the final performance. The main contributions of this work are summarized as follows.
\begin{itemize}%[leftmargin=1.5em]
	\item This paper investigates the effects of adapting different features and network layers on the ICMH performance. It reveals that different features and layers contribute unequally to downstream task performance, providing a quantitative foundation of adaptation for ICMH.
	
	\item An importance guided feature and network dual adaptation (IGDA) method is developed for ICMH. It is designed based on the low-rank adaptation and importance guided parameter allocation to improve the task performance under a parameter budget.
	
	\item A Kronecker product based adaptation method is developed for feature adaptation in order to further improve its representation efficiency.
	
\end{itemize}
Extensive experiments have been conducted on representative machine vision tasks including image classification and object detection. The results demonstrate that  the proposed framework outperforms the existing methods on all tasks.

\section{Related Work}

\subsection{Image Compression for Human Vision}
With the rapid development of deep learning, learned image compression has achieved remarkable progress, and its performance has surpassed traditional encoding standards such as JPEG. It usually takes a variational autoencoder architecture, including encoder, decoder, quantization and an entropy model \cite{ballé2018variational}. Many methods have been developed to improve the encoder/decoder transforms and entropy models. Ballé \emph{et al.} \cite{ballé2018variational} introduced a hyperprior network to  better formulate the distribution of the latent features. Minnen \emph{et al.} \cite{NEURIPS2018_53edebc5} proposed an autoregressive context model upon the hyperprior framework, leveraging already encoded neighboring latent features to predict the probability distribution of the following to-be-coded features. %Regarding the encoder/decoder transforms, Transformer \cite{vaswani2017attention}, invertible network \cite{dinh2016density} and combination of different transforms have also been explored in the existing works \cite{liu2023learned,gao2025approximately,cheng2020learned,xie2021enhanced,li2024frequencyaware}.Various entropy models and transforms have been explored in the VAE based neural image compression framework \cite{koyuncu2022contextformer,cheng2020learned,minnen2020channel,qian2022entroformer}, which are not further detailed here.
Regarding the encoder/decoder transforms, Transformer \cite{vaswani2017attention}, invertible network \cite{dinh2016density,11003432,GUO2026112532} and combination of different transforms have also been explored in the existing works \cite{liu2023learned,gao2025approximately,cheng2020learned,xie2021enhanced,li2024frequencyaware,11296866,pengligao}.Various entropy models and transforms have been explored in the VAE based neural image compression framework \cite{koyuncu2022contextformer,cheng2020learned,minnen2020channel,qian2022entroformer,jin2021deep,wang2020ensemble,sheng2024spatial,li2024high,zhang2025learning,10129217,11075899,11112672,11251299,GAO10061473,GAO10256132,GAO11419884,jiangyuntian}, which are not further detailed here.

On the other hand, a new compression paradigm based on implicit neural representations \cite{dupont2021coin,dupont2022coin,damodaran2023rqat,guo2023compression,chen2021nerv,chen2023hnerv,mildenhall2021nerf,11329185,wang2025metricgrids,ladune2026cool} has emerged in recent years. It learns a neural network to represent an image. Dupont \emph{et al.} \cite{dupont2021coin} employed a neural network to fit an image, and used the trained weights as a compressed representation. SIREN \cite{sitzmann2020implicit} introduced periodic activation functions to help represent high-frequency image details. COOL-CHIC \cite{ladune2023cool} introduced a hierarchical feature representation to capture information from low frequency to high frequency and an entropy model with quantization to improve compression efficiency of the features and model weights. Building upon COOL-CHIC, C3 \cite{kim2024c3} further introduced techniques such as soft-rounding and Kumaraswamy noise to improve the compression performance. EQLINR \cite{zhang2023enhanced} introduced an enhanced quantified local implicit neural representation method to explore convolution in the local processing and improve the quantization with a stochastic gumbel annealing strategy. These INR methods are optimized for human vision without considering the adaptation to the machine vision tasks.

\subsection{Image Compression for Machine Vision}
With the increasing machine vision applications,  more and more images are used for machine vision tasks. Therefore, image compression for both machine and human vision (ICMH) has become popular in recent years \cite{choi2022scalable,liu2021semantics,liu2023icmh,yang2021towards,chen2023transtic,feng2023semantically,fischer2022boosting,bai2022towards,agiza2024mtlora,li2025task,zhang2024all,feng2023prompt,li2024image,park2025test,li2024human,shen2024image,lu2024preprocessing,yang2024task,feng2022image,10508463,zhang2024machine,liang2026burst}.

Earlier studies \cite{choi2022scalable,liu2021semantics,liu2023icmh,yang2021towards} usually employed a scalable codec for human and machine vision. Choi \emph{et al.} \cite{choi2022scalable} proposed to partition the latent space into two layers, with the base layer carrying semantic information for machine vision and the enhancement layer carrying high-frequency visual details for human vision. There are also works \cite{feng2023semantically,bai2022towards,zhang2024all,feng2023prompt,feng2022image} that use a unified encoder feature and different decoders for multiple tasks.  Feng \emph{et al.} \cite{feng2022image} leveraged self-supervised learning and an information filtering module to learn an omni-feature that can support various downstream machine tasks. Recent studies \cite{chen2023transtic,fischer2022boosting,agiza2024mtlora,li2025task,li2024image,park2025test} increasingly employ a lightweight trainable module to adapt a pre-trained image codec optimized for human vision to various machine vision tasks. Li \emph{et al.} \cite{li2025task} proposed a learnable quantization step predictor, to dynamically allocate quantization steps for machine tasks. Li \emph{et al.} \cite{li2024image} introduced a lightweight adapter to modulate features in both spatial and frequency domains, effectively reducing task-irrelevant information. MTLoRA \cite{agiza2024mtlora} proposed a Low-Rank Adaptation (LoRA) \cite{hu2022lora} based framework with task-agnostic and task-specific low-rank modules  to balance shared and task-specific feature learning. Park \emph{et al.} \cite{park2025test} introduced a singular value decomposition‑based low‑rank adaptation (SVD-LoRA) \cite{baazy2024loraxs} which fine‑tuned only a small number of singular values in the decoder during inference. In addition to the adaptation of the features, preprocessing of the input image has also been explored \cite{shen2024image,lu2024preprocessing,yang2024task}. {Different from the above preprocessing pipelines, Liang \emph{et al.} \cite{liang2026burst} introduced burst image quality assessment to estimate the task-driven quality of individual frames in a burst sequence, showing that different frames contribute unequally to downstream tasks and enabling task-aware frame selection to reduce storage and transmission redundancy.} The above methods focus on ICMH based on a general image codec.

On the other hand, ICMH based on Implicit Neural Representations (INR) has remained underexplored. In \cite{li2024human}, Li \emph{et al.} applied INR to ICMH where the feature embedding is shared and different decoders are used for different tasks. This work also investigates the INR method for ICMH. Different from the general image codec, the number of parameters in the feature embedding and neural network in INR directly contributes to the bit rate. Therefore, this work focuses on developing a compact feature and network representation to improve the applicability of INR for ICMH.

\section{Proposed Method}
\label{sec:proposed}
\subsection{Formulation}
\label{sec:formulation}
The proposed IGDA-ICMH framework builds on the INR-based image compression paradigm. INR-based image compression usually contains a feature grid and a network as the implicit representation to reconstruct an image, which is directly trained with the image compression objective. The reconstruction process can be represented as 
\begin{equation}
	\hat{I} = f_1 \circ f_2({F_1, F_2})
	\label{eq:emc0}
\end{equation}
where $F_1$ and $F_2$ represent the multi-resolution features, also known as feature grid. $f_1$ and $f_2$ represent the different neural network layers usually constructed with convolution blocks. In practice, there are usually seven resolutions of features and four network layers. Here for simplicity and clarity of the formulation, only two-resolution features ($F_1$ and $F_2$) and two network layers ($f_1$ and $f_2$) are used.

\begin{figure*}[t] 
	\centering
	\includegraphics[width=0.9\textwidth]{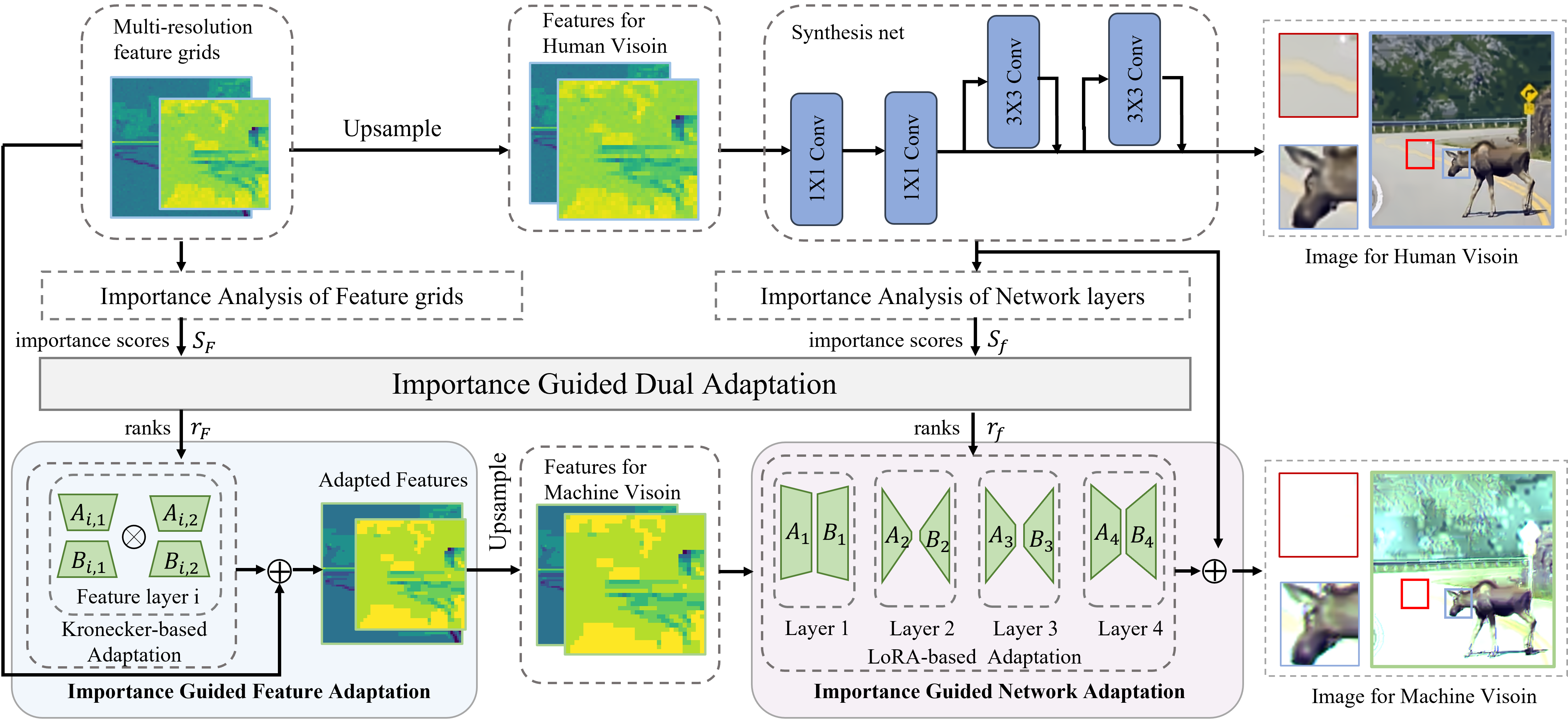} % 调整为页面宽度的80%
	%\caption{Overview of the proposed IGDA-ICMH framework, with IGFA to adapt the features of different resolutions and IGNA to adapt the different layers.}
	\caption{{Overview of the proposed IGDA-ICMH framework, with IGFA to adapt the feature grids of different resolutions and IGNA to adapt different network layers. Importance of different features and network layers are estimated and then used to guide the dual adaptation. }}
	%Built on a human vision-oriented INR backbone, the framework first analyzes the importance of the multi-resolution feature grids and network layers for machine vision tasks, yielding the importance scores $S_F$ and $S_f$, respectively. These scores then guide dual adaptation for parameter allocation, where $r_f$ and $r_F$ denote the ranks assigned to network layers and feature grids, respectively. This leads to two key designs: IGFA to adapt the features of different resolutions and IGNA to adapt different network layers.}}
	\label{fig:your_framework}	
\end{figure*}

In the task of ICMH, the reconstructed image is evaluated differently for the human and machine vision tasks, with more semantic information for machine vision tasks and general appearance for human vision. It is difficult to learn a unified INR network to directly work for both human and machine vision tasks, as evidenced in the experiments later shown in Fig. \ref{fig:comparemodule}. Therefore, we propose a feature and network dual adaptation framework for INR-based ICMH. It can be represented as
\begin{equation}
	\begin{split}
		\hat{I}_H &= \left(f_1 + \Delta f_{1,H} \right) \circ \left(f_2 + \Delta f_{2,H} \right) \\
		&\quad \left( \left[ F_1 + \Delta F_{1,H} \right], \left[ F_2 + \Delta F_{2,H} \right] \right) \\
		\hat{I}_M &= \left(f_1 + \Delta f_{1,M} \right) \circ \left(f_2 + \Delta f_{2,M} \right) \\
		&\quad \left( \left[ F_1 + \Delta F_{1,M} \right], \left[ F_2 + \Delta F_{2,M} \right] \right)
	\end{split}
	\label{eq:2}
\end{equation}
where $\hat{I}_H$ and $\hat{I}_M$ represent the reconstructed images for human vision and machine vision tasks, respectively. $\Delta F_{i,T}$ and $\Delta f_{i,T}$ represent the adaptations to the feature and network, respectively, with $i$ representing the feature resolution or network layer index and $T$ representing the task. In this way, $F_i$ and $f_i$ are learned as common basic features and network layers for both tasks, while using $\Delta F_{i,T}$ and $\Delta f_{i,T}$ to adapt  the feature and network layers to reconstruct the different images for different tasks.

In this paper, the adaptation is realized by directly changing the feature and network layer similarly to LoRA. In this way, we can simplify the above two adaptation processes into one by merging the baseline with one adaptation and compensating the other one. Considering that the human vision task usually requires more complete image information than the machine vision task, {and that the INR codec for human vision already preserves rich texture and structure information,} the INR for the human vision task is treated as the main branch and the adaptation is added for the machine vision task. {This can keep the human vision reconstruction quality unaffected while adapting the shared representation to machine vision.} Accordingly, Eq. \ref{eq:2} can be further simplified as 
\begin{equation}
	\begin{aligned}
		\hat{I}_H &= f_1 \circ f_2(F_1, F_2) \\
		\hat{I}_M &= \left(f_1 + \Delta f_1 \right) \circ \left(f_2 + \Delta f_2 \right) \\
		&\quad \left( \left[ F_1 + \Delta F_1 \right], \left[ F_2 + \Delta F_2 \right] \right)
	\end{aligned}
	\label{eq:emc}
\end{equation}
where $f_i$ and $F_i$ reconstruct the image for the human vision, $\Delta f_i = \Delta f_{i,M} - \Delta f_{i,H}$ and $\Delta F_i = \Delta F_{i,M} - \Delta F_{i,H}$ {represent the additional network and feature adaptation terms for machine vision.}
%adapt to the machine vision. 

Built on the adaptation framework with a pretrained human vision codec where the baseline provides a machine vision loss $D_0$ and rate $R_0$, {the rate-distortion optimization is then applied to the adaptation terms $\Delta F_i$ and $\Delta f_i$ in Eq. \ref{eq:emc} for machine vision}, and the adaptation process can be optimized by
\begin{equation}
	\min D_0+\Delta D \sim \min \Delta D,\quad \text{s.t.} \quad \Delta R \leq R_P
	\label{eq:deltard}
\end{equation}
where $\Delta D$ and $\Delta R$ represent the rate and distortion change in the adaptation, $R_P$ denotes the available rate budget for the adaptation.

In INR-based compression methods, rate is determined by the parameters, including both the feature grid and network parameters, and similarly $\Delta R$ is composed of the adaptation parameters. 
However, the importance of the different resolutions of features and different network layers is different between the human vision and machine vision tasks. In other words, as demonstrated in the Introduction, the improvements by adapting the different layers of a human-vision-targeted model for machine vision are quite different. This makes uniform adaptation inefficient to reduce $\Delta D$. 
%Therefore, a dual adaptation strategy is developed to assign the number of parameters for the adaptation at different resolutions of features and network layers.
Considering the different distortion change properties of different layers and features, Eq. \ref{eq:deltard} can be further expressed as 
\begin{equation}	
	\min \sum_{i=1}^{{L_F}}{\Delta D_i(a_i)}+\sum_{j=1}^{{L_f}}{\Delta D_j(a_j)}, \quad \text{s.t.} \quad \sum_{i=1}^{{L_F}}a_i +\sum_{j=1}^{{L_f}} a_j \leq P
	\label{eq:minp}
\end{equation}
where $a_i$ and $a_j$ denote the number of adaptation parameters allocated to the $i$th feature layer and $j$th network layer, respectively. {${L_F}$ and ${L_f}$ represent the number of feature layers and network layers for adaptation, respectively.} $\Delta D_i$ and $\Delta D_j$ represent the corresponding distortion change, respectively. Since in INR $\Delta R$ is proportional to the number of parameters, for simplification, the number of adaptation parameters is directly used to represent the rate change.

This reformulation clarifies that, in INR-based ICMH, the key challenge is not merely to design another adapter structure, but to determine how a limited adaptation parameter budget allocated across feature and network layers to minimize the downstream machine task distortion. To simplify the optimization process, an importance guided dual adaptation strategy is developed to assign the number of parameters for the adaptation at different resolutions of features and network layers based on the importance to the final performance. Specifically, a two-stage importance-guided allocation strategy is adopted. In the first stage, the total budget $P$ is divided into two parts, $P_f$ for network adaptation and $P_F$ for feature adaptation.  Then the parameters are further allocated in different layers of each branch. This process can be expressed as
\begin{equation}
	\begin{split}
		&\begin{aligned}
			\hat{I}_M &= \left(f_1 + T_{f_1}(S_{i,f_1})\right) \circ \left(f_2 + T_{f_2}(S_{i,f_2})\right) \\
			&\quad \left( \left[ F_1 + T_{F_1}(S_{i,F_1}) \right], \left[ F_2 + T_{F_2}(S_{i,F_2}) \right] \right)
		\end{aligned}
	\end{split}
	\label{eq:imt}
\end{equation}
where $S_i$ refers to the importance score of each feature or network layer, determining the distortion change $\Delta D_i(a_i)$ and $\Delta D_j(a_j)$. $T$ is an operator mapping the importance score to the space of the $\Delta f$ or $\Delta F$. In implementation, $T$ sets the number of parameters in each adaptation using LoRA with importance guided rank allocation as further explained in the following subsections. 
%proportional to the importance score as further explained in the following subsections. 
%and accordingly the rank used in LoRA.

%an importance lookup table on the different resolutions of features and network layers is constructed and then used to guide the adaptation. 

Based on the above formulation, the proposed IGDA-ICMH framework can be applied to any INR-based ICMH method. In this paper, the COOL-CHIC, which excels at human-oriented image reconstruction, is used as the baseline INR model. The overall network of the proposed IGDA-ICMH based on COOL-CHIC is shown in Fig. \ref{fig:your_framework}. It contains a main branch to reconstruct the image for human vision, importance guided feature adaptation (IGFA) and importance guided network adaptation (IGNA) modules. The main branch contains a feature grid with features of  seven different resolutions and four convolution blocks with different filter sizes (1x1,3x3). {As shown in Fig. 2, IGFA introduces task-specific adaptation to feature grids at different resolutions by combining low-rank and Kronecker-product forms, while IGNA performs weight adaptation on different network layers in the form of LoRA. The size of the component matrices A and B in the adaptation are determined by the importance scores of the corresponding layers. Quantization and entropy modeling are also performed on the learned parameters.}

The proposed framework can be trained in two ways. First, the whole model, containing the main branch, IGFA, and IGNA modules, can be jointly optimized with both human vision and machine vision objectives. Second, the framework can also be used to directly fine-tune a pretrained network for different tasks. In this case, the pretrained parameters of the main branch are frozen, while only the adapter modules are trainable. The rate-distortion loss is used for supervision. For different machine vision tasks, the distortion measure is also different and detailed in the experiment section. %For human vision, MSE is used, for image classification, cross-entropy is used and for object detection, the .

%The whole model, containing the main branch, IGFA, and IGNA modules, can be jointly optimized in an end-to-end manner. The framework can also be used to directly fine-tune a pretrained network for different tasks. In this case, the pretrained parameters of the main branch are frozen, while only the adapter modules are trainable. 

%The overall objective is designed based on the rate distortion cost, where the distortion is evaluated with both human vision and machine vision. It can be expressed as
%\begin{equation}
%	L = D_H + \lambda_1 D_M + \lambda_2 R
%\end{equation}
%where $D_H$ and $D_M$ represent the human vision reconstruction distortion (MSE in the experiments) and  machine vision task loss, respectively. For different machine vision task, the distortion measure is also different. For example, cross-entropy is used for the classification task. $R$ denotes the bitrates, and $\lambda_1$ and $\lambda_2$ are factors to balance between the distortion of human vision and machine vision, and between the rate and distortion, respectively.

%The framework can also be used to directly fine-tune a pretrained network for different tasks. In this case, the pretrained parameters of the main branch are frozen, while only the adapter modules are trainable. 

%\begin{figure}[!t] 
%	\centering   
%	\includegraphics[width=0.9\linewidth]{figures/two.png}  % 替换为你的图片文件名（
%	\caption{Illustration of the IGFA and IGNA modules.}
%	\label{fig:two}
%\end{figure}

In the following, the details of the proposed IGFA and IGNA modules are explained.

\subsection{Importance Guided Network Adaptation}
\label{sec:igna}
To achieve ICMH based on INR representation, an adaptation branch is developed to reconstruct different images for different tasks as illustrated in Eq. \ref{eq:imt}. Inspired by the success of Low-Rank Adaptation (LoRA) in fine-tuning large language models, in this paper, LoRA is used for the adaptation branch, in order to reduce the number of learnable parameters. Given the input feature map $F_i \in \mathbb{R}^{C_{in} \times H \times W}$, the adaptation process can be expressed as
\begin{equation}
	\begin{gathered}
		\Delta W_i = B_i(r_i) A_i(r_i) \\
		F_{i+1} = \left(W_i + \Delta W_i \right) \ast F_i + b_i
	\end{gathered}
	\label{eq:deltaw}
\end{equation}
where $W_i \in \mathbb{R}^{C_{out} \times C_{in} \times k \times k}$ is the weight of the $i$-th layer in the main branch. $A_i(r_i) \in \mathbb{R}^{(kr_i) \times (C_{in}k)}$, $B_i(r_i) \in \mathbb{R}^{(C_{out}k) \times (kr_i)}$ are the low-rank components to generate the low-rank adaptation weight, and  $r_i$ is the rank used. In the proposed method, the original weights of the main branch are kept intact, while only a lightweight task-specific weight $\Delta W_i$ is introduced. As a result, the human-oriented reconstruction capability learned by the pretrained INR model can be preserved, and the network can still be flexibly adapted to different downstream machine vision tasks with only a small number of adaptation parameters.

%Specifically, the importance is used to guide the setting of rank $r_i$ in order to control the number of parameters for each network layer. 
According to Eqs.~\ref{eq:minp} and \ref{eq:imt}, layers with larger importance score $S_i$ are supposed to receive larger adaptation capacity under the parameter budget. In IGNA, this adaptation capacity is instantiated by the LoRA rank. Assuming the parameter cost is $c_i$ when the rank is 1 for the $i$th layer, the parameter allocation is converted to the rank allocation as

% and denoting the parameter budget for the network adaptation branch by $P_f$, the layer importance score $S_i$, obtained from the unit parameter performance improvement measured in the importance analysis experiment as mentioned in the Introduction
%\begin{equation}
%	\min L,\quad \text{s.t.} \quad \sum_{i=1}^{N} c_i r_i \leq P
%\end{equation}
%where $L$ is the loss function and $N$ is the number of layers.

\begin{equation}
	r_i = \left\lfloor \frac{S_i}{\sum_{j=1}^{m} S_j} \frac{P_f}{c_i} \right\rfloor
\end{equation}
where $\lfloor \cdot \rfloor$ is the floor operation. $P_f$ represents the parameter budget for the network adaptation branch, the layer importance score $S_i$ is obtained  from the importance analysis experiments described in the Introduction. %Specifically, each feature grid or network layer is progressively trained while all other layers are kept frozen. The score is computed as the ratio between the downstream task performance improvement and the number of trainable parameters in that layer. Therefore, $S_i$ reflects the benefit of allocating adaptive capacity to the corresponding layer under a limited parameter budget.
%from the unit parameter performance improvement measured in the importance analysis experiment as mentioned in the Introduction.

To maintain structural compactness in the adaptation, the rank is further clipped in a range $[r_{\min}, r_{\max}]$. In extreme low‑budget scenarios, the above allocation may lead to all layers receiving $r_i < r_{\min}$, insufficient to instantiate an adaptation module. To avoid degenerate adaptation, IGNA introduces a selective adaptation strategy, where only the top‑K layers with the highest importance scores are enabled for adaptation. {Specifically, when the rank obtained by proportional allocation falls below the minimum rank required for effective training, we avoid spreading the limited budget over too many layers. Instead, we select the layer with the highest K importance scores.} This ensures that the important layers can be sufficiently adapted.  With the rank of each adaptation layer set, the adaptation process in Eq. \ref{eq:deltaw} can be performed in order to achieve efficient image compression for different tasks.

\subsection{Importance Guided Feature Adaptation}
\label{sec:igfa}
Although the feature grid provides a compact representation in multi-resolution space, the contributions of features at different resolutions also vary significantly to downstream machine vision tasks. Similar to the IGNA, an importance guided feature adaptation (IGFA) is also developed to adapt the features at different resolutions. Different from the network layer adaptation where the size of the weight in each layer is usually not large, the feature can be of very large resolution, over $800 \times 800$. In this case, even LoRA cannot produce an efficient adaptation with a small number of parameters. Therefore, a joint Kronecker product and low-rank representation based adaptation is developed.

 For a given feature input  $F_i \in \mathbb{R}^{H \times W}$ at the $i$th resolution, the adaptation takes a Kronecker product form to factorize the adaptation weight into two subspaces, and the two subspace matrices are further decomposed with low-rank approximation. This process can be represented as
\begin{equation}
	\Delta F_i = (B_{i,1} A_{i,1}) \otimes (B_{i,2} A_{i,2}) 
\end{equation}
where $\otimes$ represents the Kronecker product. $A_{i,1} \in \mathbb{R}^{r_{i,1} \times w_1}$, $A_{i,2} \in \mathbb{R}^{r_{i,2} \times w_2}$, $B_{i,1} \in \mathbb{R}^{h_1 \times r_{i,1}}$, $B_{i,2} \in \mathbb{R}^{h_2 \times r_{i,2}}$ are the component matrices. $h_1$, $h_2$, $w_1$ and $w_2$ represent the scales of sub-matrices in the Kronecker product and satisfy the condition $h_1 \times h_2 = H$, $w_1 \times w_2 = W$.  $r_{i,1}$ and $r_{i,2}$ denote the ranks of two low rank subspaces.%, satisfying the condition $r_{i,1} \times r_{i,2} = r_i$. 
%Then to further improve parameter efficiency, each Kronecker factor is optionally decomposed using a low‑rank representation
%\begin{equation}
%	\begin{aligned}
	%		A_{i,k} &= A_{i,k,B} A_{i,k,A}, \\
	%		B_{i,k} &= B_{i,k,B} B_{i,k,A}, \\
	%		k &\in \{1, 2\}
	%	\end{aligned}
%\end{equation}
%Where $A_{i,k,B} \in \mathbb{R}^{r_{i,k} \times r_{inner}}$, $A_{i,k,A} \in \mathbb{R}^{r_{inner} \times H_k}$, $B_{i,k,B} \in \mathbb{R}^{H_k \times r_{inner}}$, $B_{i,k,A} \in \mathbb{R}^{r_{inner} \times r_{i,k}}$. $A_{i,k,B}$ and $A_{i,k,A}$ represent the matrices after low rank decomposition of$A_{i,k}$, $B_{i,k,B}$ and $B_{i,k,A}$ represent the matrices after low rank decomposition of $B_{i,k}$, $r_inner$ denotes the rank of two low rank subspaces.

This decomposition provides a compact yet expressive structure by exploring the local properties of the image. The parameters $H_i$ and $r_{i}$ in the Kronecker product  and low-rank representation are determined by the corresponding importance score $S_i$, in the same way as the IGNA. 
%In addition, the number of adaptation parameters between IGNA and IGFA \replaced[id=R3]{is}{are} also determined by evaluating their effects \replaced[id=R3]{on}{to} final performance. %Specifically, the network layers and features are independently optimized, estimating the average improvement in accuracy per parameter for both layers and features. Then similarly as in the IGNA and IGFA, different parameter budgets are allocated for the network layers and features. Through this strategy, the framework ensures parameter resources to be concentrated in areas with higher performance benefits.

\section{Experiment}
\subsection{Experimental Setting}

\textbf{ICMH tasks and datasets}. To comprehensively evaluate the proposed framework, extensive experiments are conducted on {three representative machine vision tasks, including image classification, object detection and instance segmentation}. Four widely used datasets are adopted for evaluation, including the Oxford 102 Flowers dataset \cite{xia2017inception} and Animals with Attributes 2 (AwA2) dataset \cite{xian2017zero} for image classification, {CrowdHuman dataset \cite{shao2018crowdhuman} for object detection and COCO2017 dataset \cite{lin2014microsoft} for object detection and instance segmentation}. 

\begin{figure*}[t]
	\centering
	
	% 第一行 纯图片
	\subfloat[Image Classification on Oxford 102 Flowers.]{
	\includegraphics[width=0.420\textwidth]{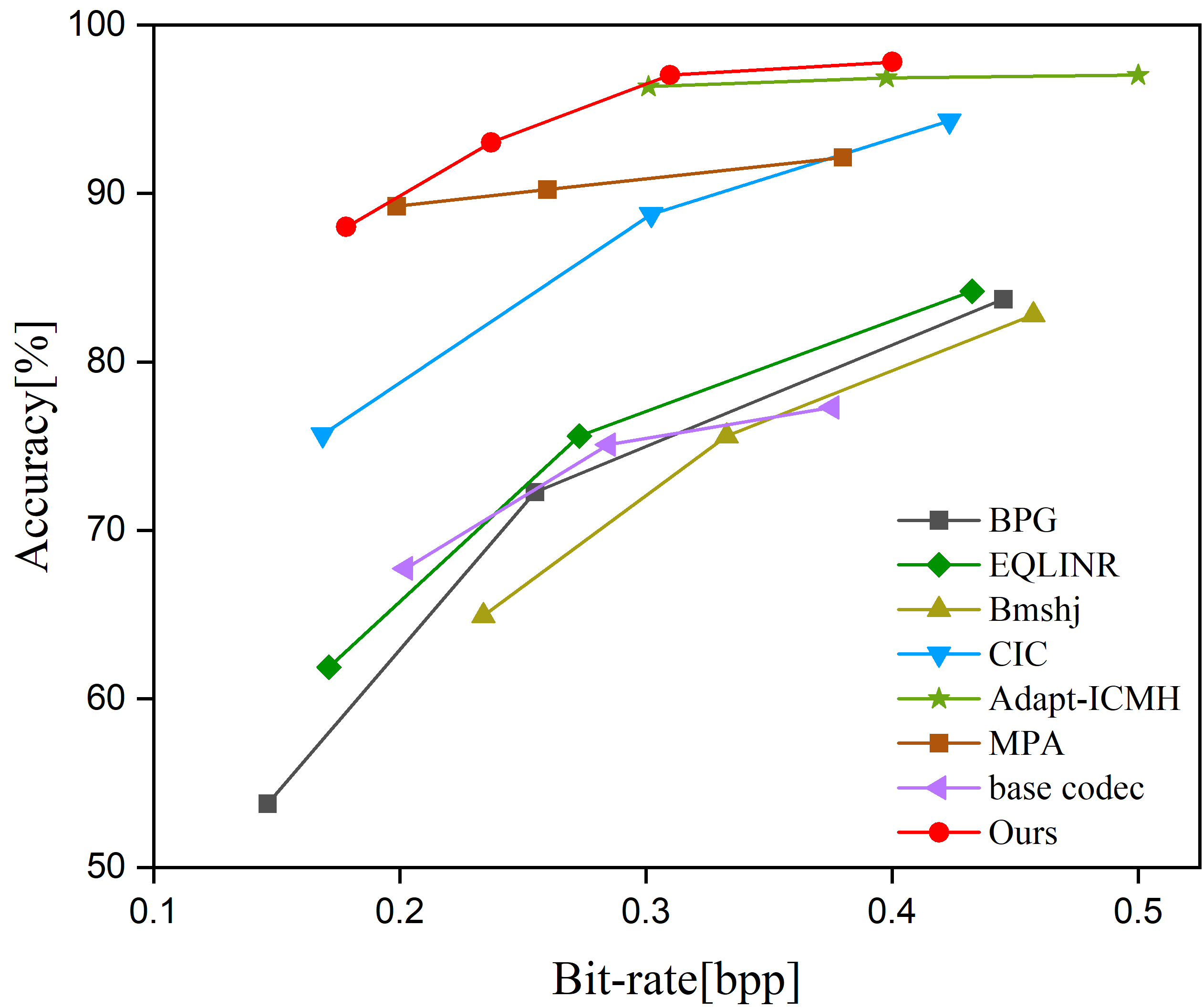}%
	\label{fig:r-a-flower}
	}
	\hspace{0.08\textwidth}
	\subfloat[Image Classification on Animals with Attributes 2.]{
	\includegraphics[width=0.420\textwidth]{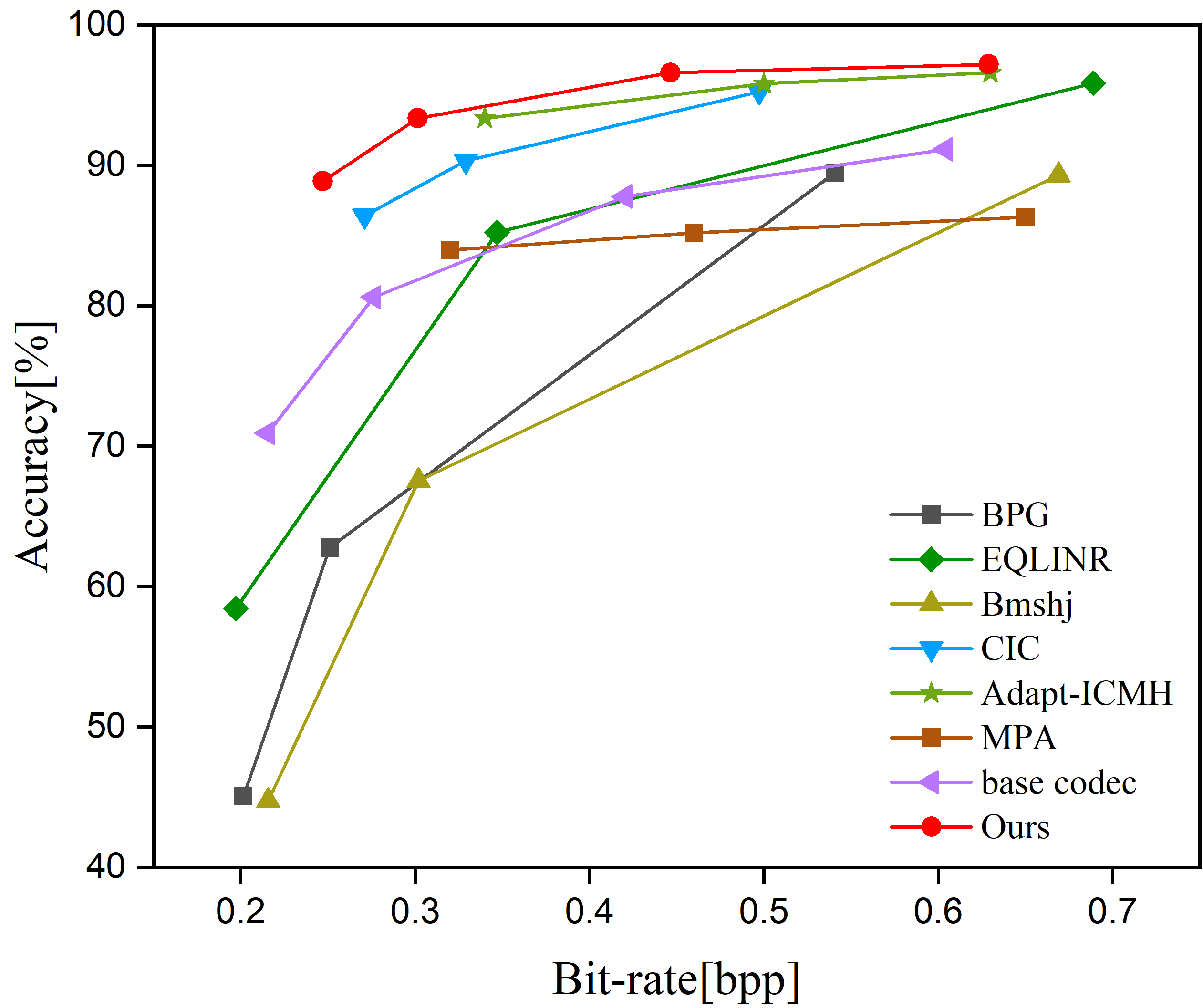}%
	\label{fig:r-a-awa}
	}
	
	% 第二行
	%\par%\vspace{0.5em}
	%
	\subfloat[Object detection on CrowdHuman dataset.]{
	\includegraphics[width=0.42\textwidth]{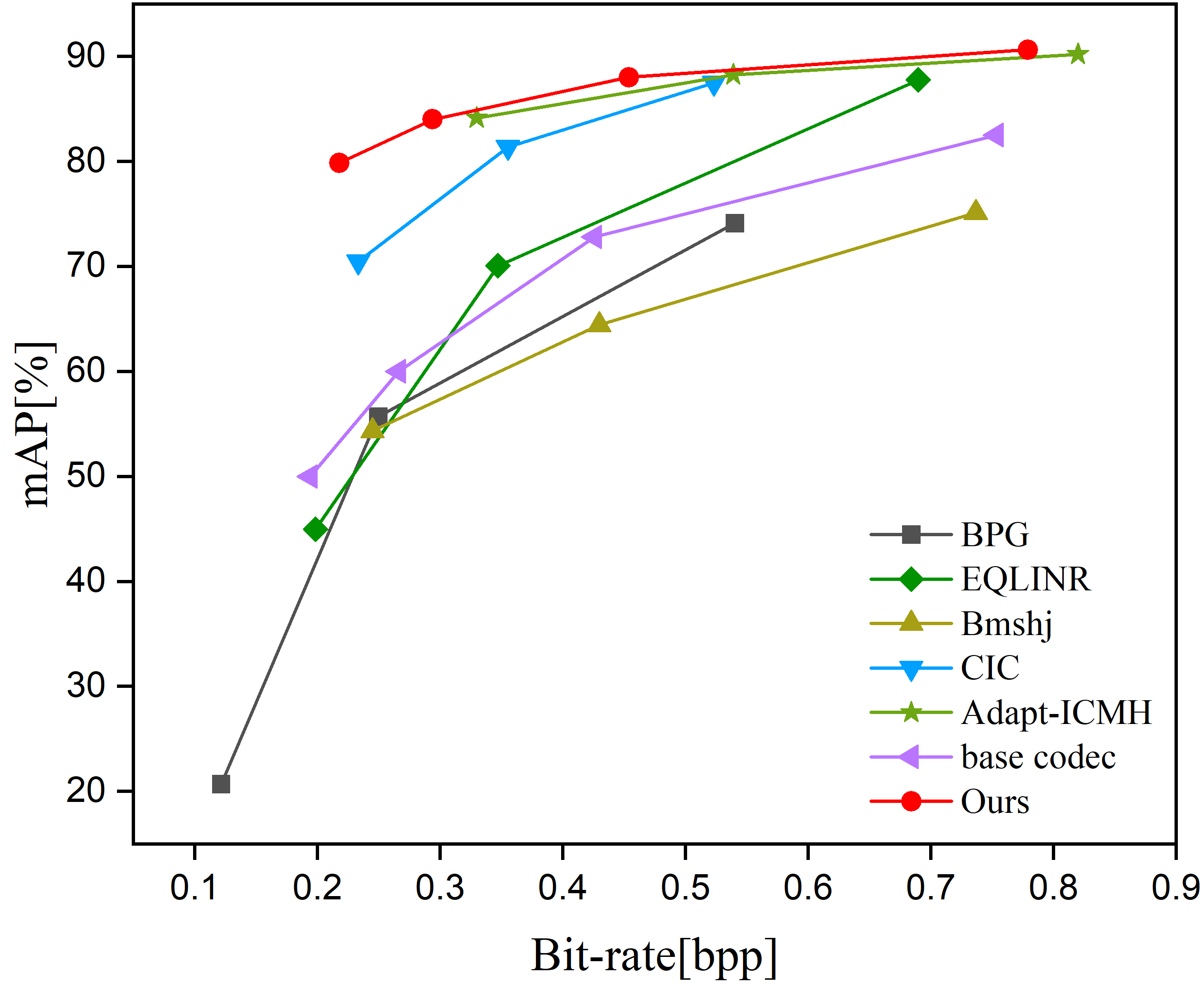}%}
	\label{fig:r-a-crowd}}	
	\hspace{0.08\textwidth}
	\subfloat[Object detection on COCO2017 datdaset.]{
	\includegraphics[width=0.42\textwidth]{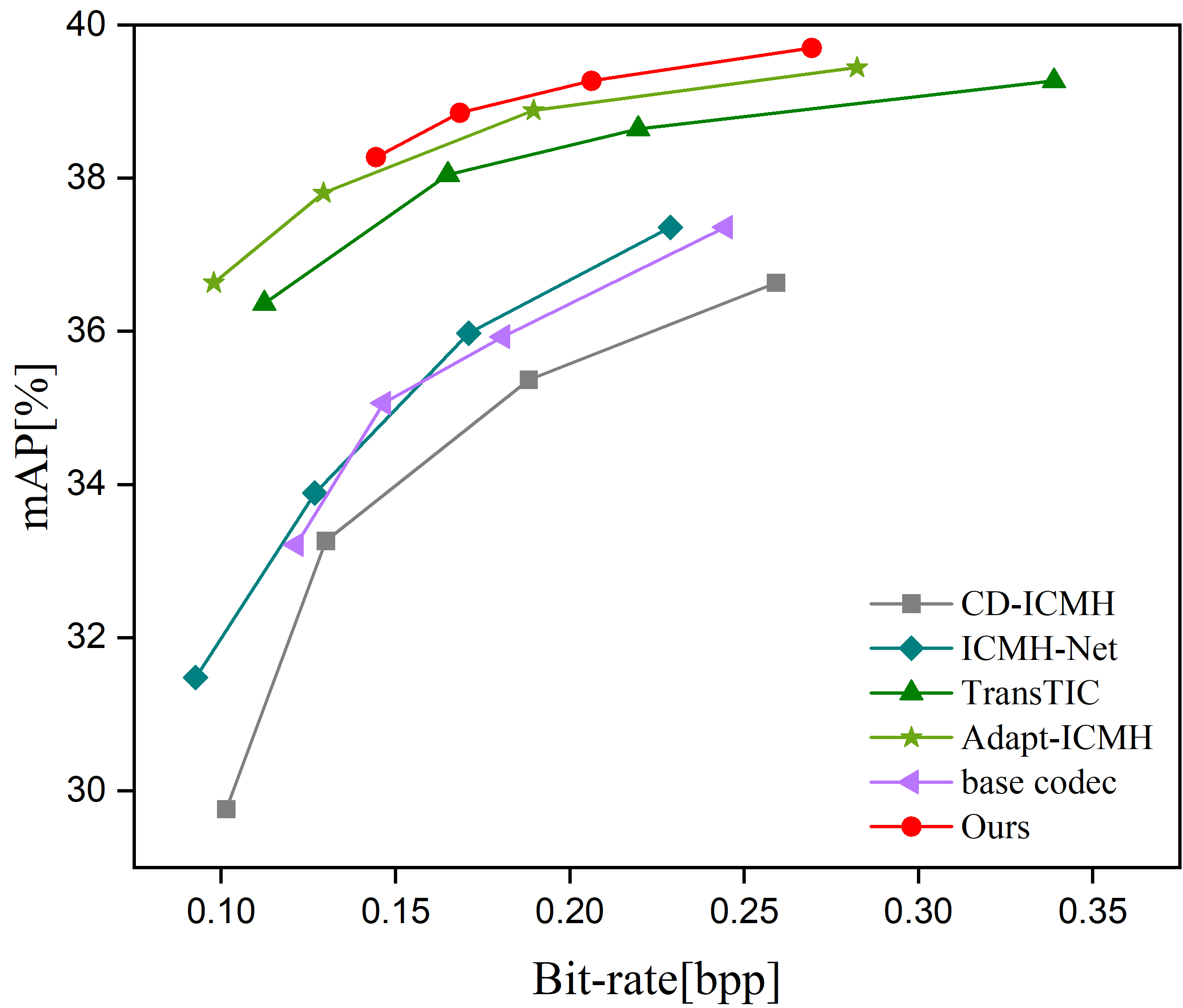}%
	\label{fig:r-a-coco}
	}
	
	\caption{Rate-Accuracy performance comparison under different machine vision tasks including image classification and object detection on different datasets.}
	\label{fig:comparemodule}
\end{figure*} 

\begin{figure*}[t]
	\centering
	
	\subfloat[]{
		\includegraphics[width=0.42\textwidth]{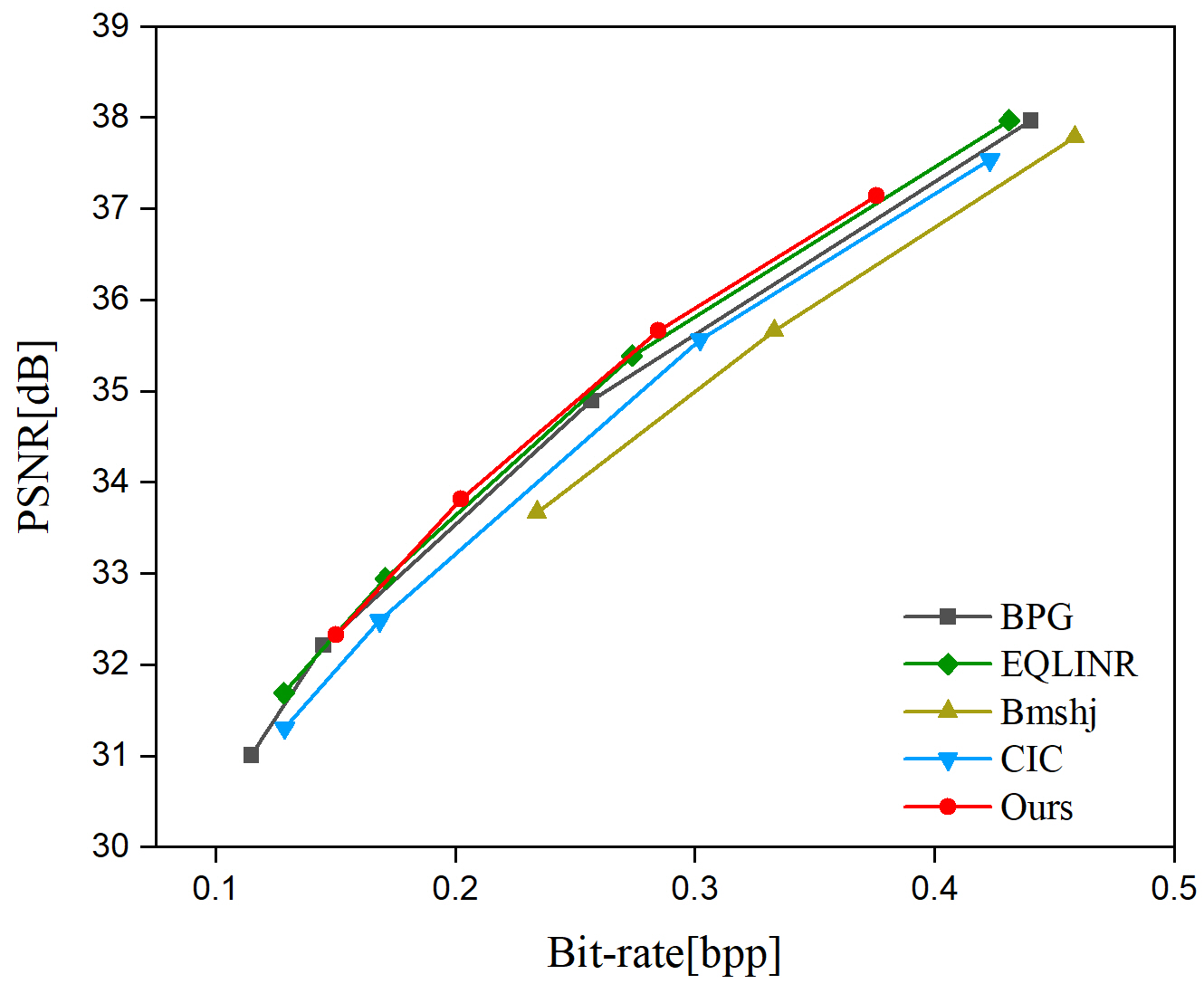}
		\label{fig:sub1}
	}%
	\hspace{0.07\textwidth}
	\subfloat[]{
		\includegraphics[width=0.42\textwidth]{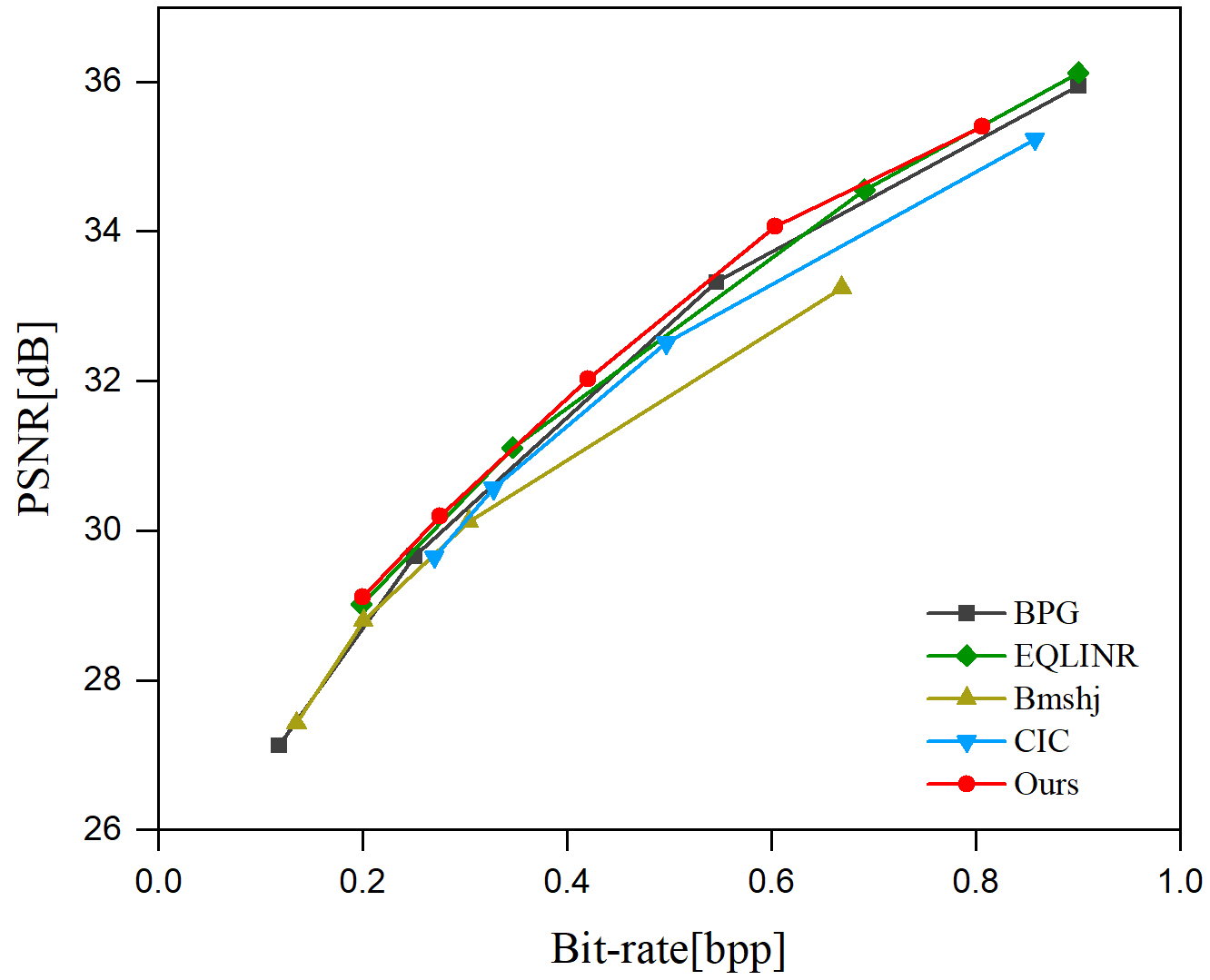}
		\label{fig:sub2}
	}
		
	\caption{Rate-distortion curves on (a) Oxford 102 Flowers and (b) Animals with Attributes 2, respectively.}
	\label{fig:three_images}
\end{figure*}
%Specifically, the Oxford 102 Flowers Dataset is a fine-grained image classification benchmark. It comprises 8,189 images spanning 102 flower categories, with significant intra-class variation and diverse visual scenes, making it suitable for evaluating the robustness of machine vision models under compressed representations.The Animals with Attributes 2 dataset is an image dataset used for animal classification, consisting of 37,322 images from 50 animal categories. The CrowdHuman dataset is widely used for object detection. It contains approximately 15,000 images and over 330,000 target instances annotated, with varying degrees of overlap and occlusion between the targets. The data covers rich outdoor and crowded environments, which can effectively support object detection task. The COCO 2017 Dataset is a large-scale dataset for dense prediction tasks, including object detection and instance segmentation. It contains 118,000 training images and 5,000 validation images across 80 object categories.

\textbf{Training and evaluation details}. 
For image classification, the Swin Transformer model \cite{liu2021swin} pre-trained on the ImageNet is used as the basic classifier. For object detection on CrowdHuman, the Detection Transformer model \cite{li2024human} is used as the basic detector while for COCO2017, the Faster R-CNN \cite{ren2015faster} is used, to be aligned with the setting in \cite{li2024human, li2024image}. {For instance segmentation on COCO2017, the Mask R-CNN \cite{he2017mask} is used as the basic segmentation model.} {The object detection and instance segmentation models are pre-trained on COCO2017.} Then the models are fine-tuned on the different datasets in the same way as in \cite{li2024human}. Finally the models are fixed to evaluate the reconstructed image produced by different methods to produce the top-1 accuracy and Average Precision (AP) for image classification{, object detection, and instance segmentation}, respectively.

The compression performance is evaluated using the rate-quality metrics, where the rate is evaluated with bits per pixel (bpp), and quality is evaluated with top-1 accuracy for the classification task and Average Precision for the object detection task {and the instance segmentation task}.  Different models are trained under different rates with the Lagrangian multipliers set to \{0.001, 0.002, 0.005, 0.010\}. BD-Rate saving and rate-distortion curves are used for performance comparison among different methods.%PSNR for the human vision task,
%-------------------------------------------------------------------------
\subsection{Performance Comparison with State-of-the-art Methods}

\begin{table}[!t]
	\centering
	\caption{Performance comparison in terms of BD-rate  and BD-mAP on COCO2017 dataset.}
	\label{tab:bdperformance}
	\renewcommand{\arraystretch}{1.2}
	\begin{tabular*}{\columnwidth}{@{\extracolsep{\fill}}>{\centering\arraybackslash}p{3cm}|cc}
		\hline
		\multicolumn{1}{c|}{Method} & BD-rate(\%) & BD-mAP(\%) \\
		\hline
		COOL-CHIC \cite{ladune2023cool} & 0 & 0 \\
		CD-ICMH \cite{liu2022improving} & 11.97 & -0.84 \\
		ICMH-net \cite{liu2022improving} & -2.39 & 0.14 \\
		TransTIC \cite{chen2023transtic} & -43.31 & 2.85 \\
		Adapt-ICMH \cite{li2024image} & -52.97 & 2.97 \\
		%Ours & -65.50 & 3.20 \\
		\textbf{Ours} & \textbf{-65.50} & \textbf{3.20} \\
		\hline
	\end{tabular*}
\end{table}

The performance of the proposed method is evaluated and compared with the existing methods including COOL-CHIC \cite{ladune2023cool}, EQLINR \cite{zhang2023enhanced}, {Bmshj \cite{ballé2018variational}, }CIC \cite{li2024human}, TransTIC \cite{chen2023transtic}, ICMH-Net \cite{liu2023icmh}, CD-ICMH \cite{liu2022improving}, Adapt-ICMH \cite{li2024image} and MPA \cite{NEURIPS2024_8395fdf3}. {The results of the existing methods such as Adapt-ICMH and ICMH-Net reported in their corresponding paper are used, where both methods use Lu2022-TIC \cite{lu2022tic} as the backbone codec.} The compared methods can be categorized into three types. The first type consists of deep learning-based image compression methods such as Bmshj \cite{ballé2018variational}. The second type consists of Implicit Neural Representation (INR)-based image compression methods, including COOL-CHIC \cite{ladune2023cool} and EQLINR \cite{zhang2023enhanced}. The third type consists of ICMH methods, including the CIC \cite{li2024human} based on INR, and TransTIC \cite{chen2023transtic}, ICMH-Net \cite{liu2023icmh}, CD-ICMH \cite{liu2022improving}{, Adapt-ICMH \cite{li2024image}, MPA \cite{NEURIPS2024_8395fdf3}} based on general training. The result comparisons on different datasets are slightly different to be aligned with the compared methods since results of some methods are not reported for certain datasets. For those with publicly available  models, the results are obtained using their models.%code and pretrained
The proposed method can be trained in two ways as described in Section \ref{sec:proposed}, including joint training with both human vision and machine vision tasks, and pre-training on human vision with the backbone then only fine-tuning the proposed IGDA modules. In the experiments, it is found that the results are similar, and thus for simplicity, here we only present the results with the pre-training and fine-tuning strategy. 
%The results for human vision in terms of rate-distortion are very similar as the compared methods. This is achieved by the backbone COOL-CHIC \cite{ladune2023cool}, and the results are shown in the supplementary material.

For image classification, the results are shown in Figs. \ref{fig:r-a-flower} and \ref{fig:r-a-awa} for the Oxford 102 Flowers and AwA2 datasets, respectively. It can be observed that  our method outperforms all compared methods. Especially comparing with the INR-based ICMH method CIC, our method shows a significant performance improvement, validating the effectiveness of the proposed IGDA method. Specifically, comparing with the baseline COOL-CHIC, our method provides a BD-Rate saving of 99.23\% and 37.59\% on the two datasets, respectively.  {It is worth noting that the ICMH methods such as TransTIC and Adapt-ICMH use relatively large networks that need to be deployed in the user side, while the proposed method is an INR-based image compression method with networks self-contained in the bitstream, similar to CIC and COOL-CHIC. Therefore, these methods target different compression paradigms and application scenarios. Accordingly, the most direct comparison for our method is with INR-based methods, and compared with CIC and COOL-CHIC, our method demonstrates significantly better performance.}
For object detection, the results are shown in Figs. \ref{fig:r-a-crowd} and \ref{fig:r-a-coco}. It can be seen that the rate-AP performance of our method is also better than the existing methods, demonstrating the effectiveness of our method. 
{For instance segmentation, the results are shown in the supplementary material, where our method consistently outperforms the backbone.}

The results for human vision in terms of rate-distortion performance are very similar to the compared methods. This is achieved by the backbone COOL-CHIC \cite{ladune2023cool}, and the results are shown in Fig. \ref{fig:three_images}. It can be observed that our method performs only slightly better than the existing methods. Due to the lack of open-source CIC and EQLINR \cite{zhang2023enhanced}, their rate-distortion curves are only available on the Oxford 102 Flowers and Animals with Attributes datasets, reported in their papers. Consequently, comparisons with CIC and EQLINR on CrowdHuman dataset and COCO2017 dataset are not included. The similar performance for human vision further verifies that the coding gains on the vision tasks including image classification and object detection are from the proposed IGDA method instead of the  base codec.

%image classification performance, object detection performance, and image reconstruction performance. 

% 导言区需加载 subcaption 宏包（已确认使用子图，必须保留）

The results in terms of BD-rate and BD-mAP are also calculated and compared with the other methods using the base codec as the baseline. Here the object detection results on the COCO2017 dataset are used. Due to the lack of open-source CIC framework, the BD-rate and BD-Accuracy cannot be calculated and thus omitted for comparison. As shown in Table \ref{tab:bdperformance}, our method achieves a significant performance improvement compared to the baseline method and the other compared methods. Our method achieved a rate reduction of -65.50\%, indicating that the proposed method has a stronger advantage in compression efficiency. %Meanwhile, our method achieved an improvement of 3.203, indicating that the downstream detection performance of the model is more fully guaranteed while maintaining the compression rate.
%Comparing with the baseline COOL-CHIC, our method provides a BD-Rate saving of 64.85\% and 65.50\%.

\begin{table}[htbp]
	\centering
	\caption{{Comparison of decoding speed in terms of FPS on Oxford 102 Flowers dataset.}}
	\label{tab:deperformance}
	\renewcommand{\arraystretch}{1.2}
	\begin{tabular*}{\columnwidth}{@{\extracolsep{\fill}}>{\centering\arraybackslash}p{3cm}|cc}
		\hline
		\multicolumn{1}{c|}{Method} & Decoding Speed \\
		\hline
		%COOL-CHIC \cite{ladune2023cool} & 6.69 \\
		%Adapt-ICMH \cite{li2024image} & 26.34 \\
		%MPA \cite{NEURIPS2024_8395fdf3} & 10.42 \\
		%Ours & 10.58 \\
		COOL-CHIC \cite{ladune2023cool}   & 149.48 \\
		Adapt-ICMH \cite{li2024image}     & 37.97  \\
		MPA \cite{NEURIPS2024_8395fdf3}   & 95.97  \\
		Ours                              & 94.52  \\
		\hline
	\end{tabular*}
\end{table}

\begin{table}[htbp]
	{
	\centering
	\caption{{Comparison of encoding complexity on the Oxford 102 Flowers. Param. count denotes number of total parameters.}}
	\label{tab:enperformance}
	%\renewcommand{\arraystretch}{1.2}
	% 将 p{xxcm} 全部替换为 m{xxcm}，m=垂直居中
	\begin{tabular*}{\columnwidth}{@{}>{\centering\arraybackslash}m{2.2cm}|>{\centering\arraybackslash}m{1.2cm}>{\centering\arraybackslash}m{1.1cm}>{\centering\arraybackslash}m{1.1cm}>{\centering\arraybackslash}m{1.4cm}}
		\hline
		Method & MAC/pixel & Encoding time(s) & Param. count & Trainable param \\
		\hline
		COOL-CHIC \cite{ladune2023cool} & 1435 & 202.77 & 88713 & 88713 \\
		Ours & 2340 & 86.07 & 89686 & 973 (1.1\%) \\
		\hline
	\end{tabular*}
    }
\end{table}

\subsection{Complexity Comparison}
The complexity of the proposed method is also evaluated and compared with the existing methods. For compression for human vision, the proposed method becomes the base codec COOL-CHIC, and thus only the compression branch for machine vision task is needed. For fair comparison, COOL-CHIC, Adapt-ICMH \cite{li2024image} and MPA \cite{NEURIPS2024_8395fdf3} methods with available source are used and all methods are run on the same GPU. The complexity is measured in terms of the decoding speed with FPS (frame per second) and the image compression task is used. The results are presented in Table \ref{tab:deperformance}. It can be observed that the decoding speed of our method is similar to the baseline and MPA, and faster than Adapt-ICMH. Meanwhile, as shown in Fig. \ref{fig:comparemodule}, our method attains higher image classification accuracy than both methods, validating its effectiveness. {Table \ref{tab:enperformance} further presents the encoding computational complexity comparison with the baseline. It can be observed that our method is a lightweight framework, and achieves performance improvement with a relatively low additional overhead.}

\begin{table}[!t]
	\centering
	\caption{Ablation results of the proposed IGNA.}
	\label{tab:ignaperformance}
	\renewcommand{\arraystretch}{1.2}  % 与参考表格保持一致的行间距
	% 完全复刻参考表格的tabular*格式，第一列固定宽度+居中+竖线分隔
	\begin{tabular*}{\columnwidth}{@{\extracolsep{\fill}}>{\centering\arraybackslash}p{2cm}|ccc}
		\hline  % 替换booktabs的\toprule为\hline，避免竖线截断
		\multicolumn{1}{c|}{Method} & {Trainable param} & Bpp & Accuracy (\%) \\
		\hline  % 替换booktabs的\midrule为\hline
		baseline & - & 0.202 & 67.74 \\
		\hline
		%$L_1$(r=2) & 110 & 0.209 & 90.58 \\
		$L_2$(r=2) & 102 & 0.209 & 89.29 \\
		$L_3$(r=2) & 108 & 0.209 & 87.74 \\
		$L_4$(r=2) & 108 & 0.209 & 89.41 \\
		\hline
		SL(r=2) & 110 & 0.209 & 90.58 \\
		SL(r=4) & 220 & 0.213 & 91.61 \\
		SL(r=8) & 440 & 0.219 & 91.48 \\
		\hline
		AL(r=2) & 428 & 0.219 & 90.70 \\
		\hline
		% 核心修改：用multirow合并3行，设置居中显示IGNA，其余两行该列留空
		\multirow{3}{*}{\centering IGNA} & 165 & 0.210 & 90.96 \\
		& 218 & 0.213 & 91.87 \\
		& 537 & 0.222 & 92.00 \\
		\hline  % 替换booktabs的\bottomrule为\hline
	\end{tabular*}
	%\vspace{-0.6mm}
\end{table}

%-------------------------------------------------------------------------
\subsection{Ablation Studies}

Ablation studies are conducted to validate the effectiveness of the proposed IGNA and IGFA modules. The pre-trained COOL-CHIC is used as the baseline. Image classification is used as the machine vision task.

\textbf{Evaluation on the IGNA.} The effectiveness of the proposed IGNA is first evaluated. Experiments are conducted under the Lagrangian multiplier 0.005. IGNA is compared with the baseline, the $i$th layer($L_i$) adaptation, all-layer average (AL) adaptation and the single most important layer (SL) adaptation variants under various parameter budgets. Note that $L_i$ refers to adapting the $i$th layer (excluding the most important one), SL refers to adapting the most important layer according to our importance analysis, and AL refers to adapting all layers without conducting any importance-based selection.
The results  are shown in Table \ref{tab:ignaperformance}. 
  {\textit{\textbf{Firstly}}}, by comparing with the baseline, it can be seen that the proposed IGNA significantly improves the performance over the baseline with only a slight increase of the parameters and bits. With around 5\% bit increase, the performance is improved to 90.96\% from 67.74\%, in terms of accuracy.  
  {\textit{\textbf{Secondly}}}, SL(r=2), adapting the singly most important layer, outperforms all $L_i$(r=2) settings, adapting other layers. This demonstrates that our importance analysis successfully identifies the most effective layer for adaptation.   
  {\textit{\textbf{Thirdly}}}, by comparing with the SL variants and the AL(r=2), it can be seen that the proposed IGNA can achieve better performance than both adaptation settings. Especially with the number of parameters increasing, the SL variants begin to lower the performance when the rank of LoRA is larger than 4, while our method continues improving the performance. This makes sense since for the SL variants the parameters are only allocated to the layer with the highest importance score and its performance is better under limited parameter budget. When the parameter budget gets larger, its performance improvement saturates. On the other hand, our IGNA can allocate the parameters to different layers according to their importance scores and thus achieves better performance. Note that the backbone uses quantization optimization for the feature and network parameters, thus the bitrate is only slightly increased even with a relatively large increase of the parameter quantities. This is also the reason that with the number of trainable parameters further increasing, the performance of the SL variant begins to decrease.

\begin{table}[!t]
	\centering
	\caption{Ablation results of the proposed IGFA based on Kronecker product.}
	\label{tab:igfaperformance}
	\renewcommand{\arraystretch}{1.2}
	\begin{tabular*}{\columnwidth}{@{\extracolsep{\fill}}>{\centering\arraybackslash}p{2cm}|ccc}
		\hline
		\multicolumn{1}{c|}{Method} & {Trainable param} & Bpp & Accuracy (\%) \\
		\hline
		baseline & - & 0.202 & 67.74 \\
		\hline
		$L_1$(r=1) & 512 & 0.221 & 91.09 \\
		$L_2$(r=2) & 512 & 0.221 & 90.96 \\
		$L_3$(r=2) & 256 & 0.214 & 91.09 \\
		%$L_4$(r=2) & 128 & 0.209 & 91.22 \\
		$L_5$(r=2) & 64 & 0.205 & 67.74 \\
		$L_6$(r=2) & 32 & 0.203 & 67.74 \\
		$L_7$(r=2) & 16 & 0.202 & 67.74 \\
		\hline
		SL(r=2) & 128 & 0.209 & 91.22 \\
		SL(r=4) & 256 & 0.214 & 91.74 \\
		SL(r=8) & 512 & 0.221 & 91.35 \\
		\hline
		AL(r=1) & 960 & 0.239 & 91.35 \\
		\hline
		\multirow{3}{*}{\centering IGFA-L} & 448 & 0.219 & 91.48 \\
		& 960 & 0.239 & 92.12 \\
		& 2560 & 0.292 & 92.38 \\
		%& 4032 & 0.368 & 92.51 \\
		\hline
		\multirow{3}{*}{IGFA-K} & 188 & 0.213 & 91.22 \\
		& 348 & 0.217 & 91.87 \\
		& 436 & 0.219 & 92.00 \\
		\hline
	\end{tabular*}
	%\vspace{-0.6em}
\end{table}

\begin{figure*}[!t]
	\centering
	\subfloat{%
		\includegraphics[width=0.5\textwidth]{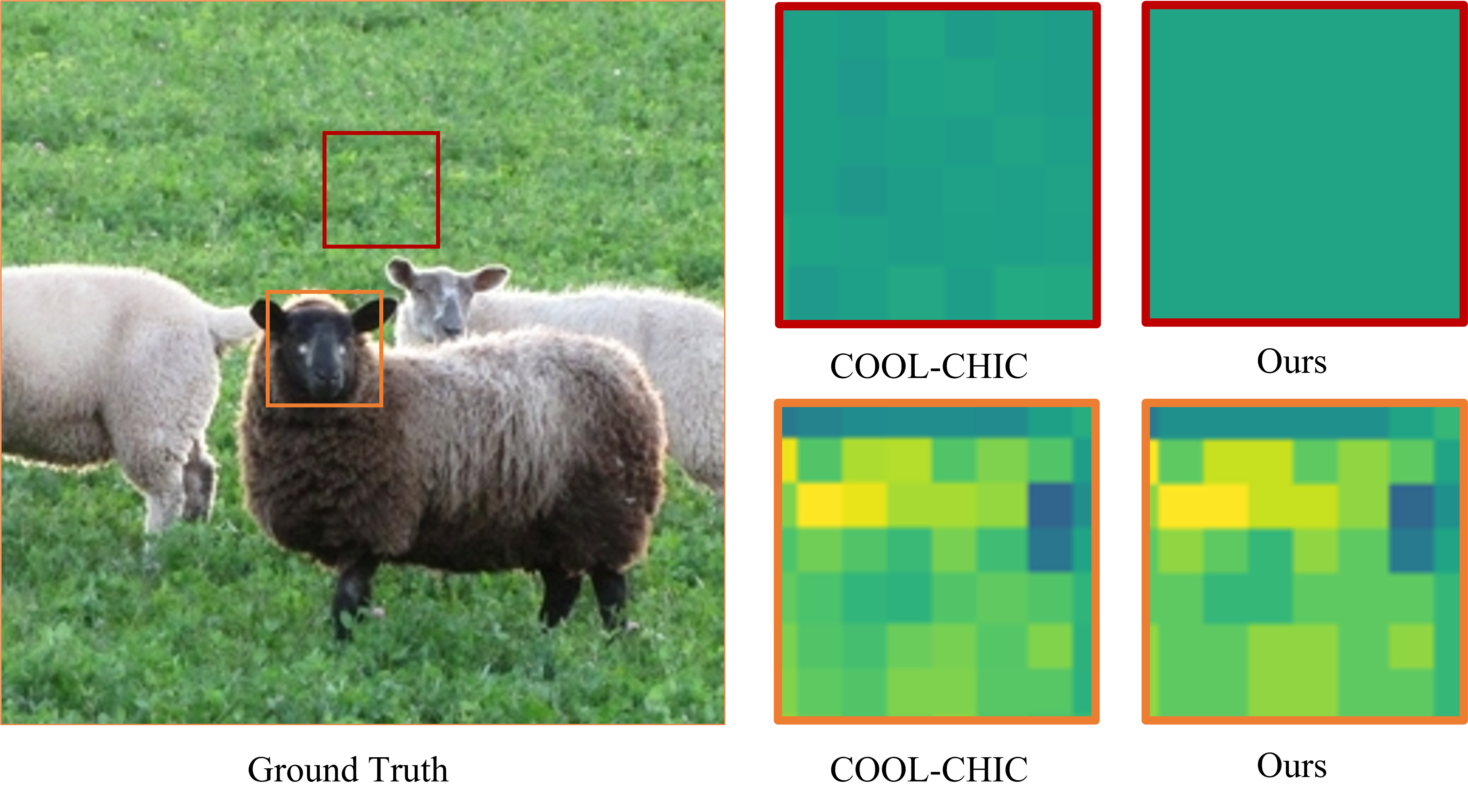}%
		\label{fig:visual10}
	}% 关键：删除换行，用%注释掉行尾空白符
	\subfloat{%
		\includegraphics[width=0.5\textwidth]{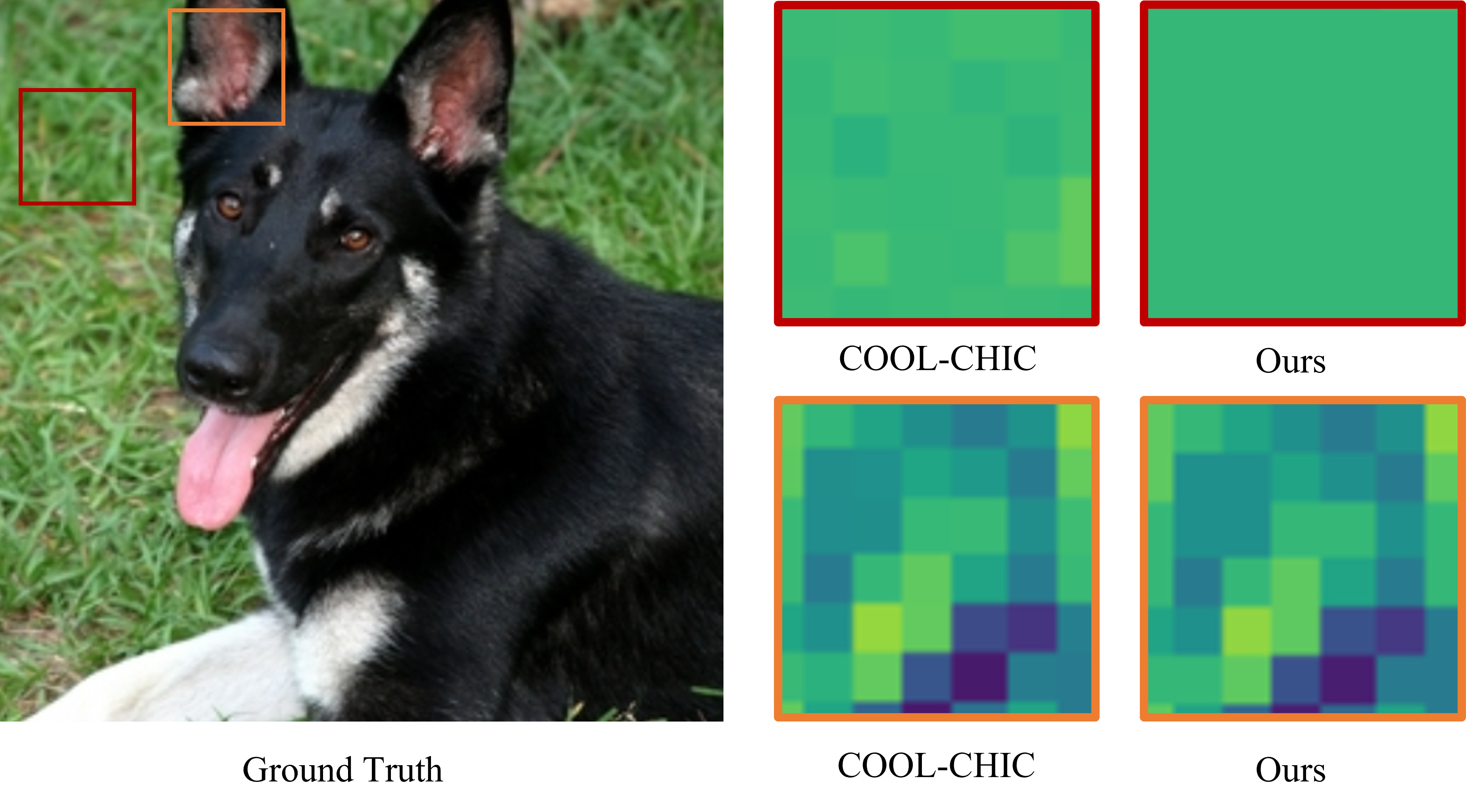}%
		\label{fig:visual12}
	}
	\par\medskip  % 换行并添加小间距

	%\par\vspace{0.6ex}
	\subfloat{%
		\includegraphics[width=0.5\textwidth]{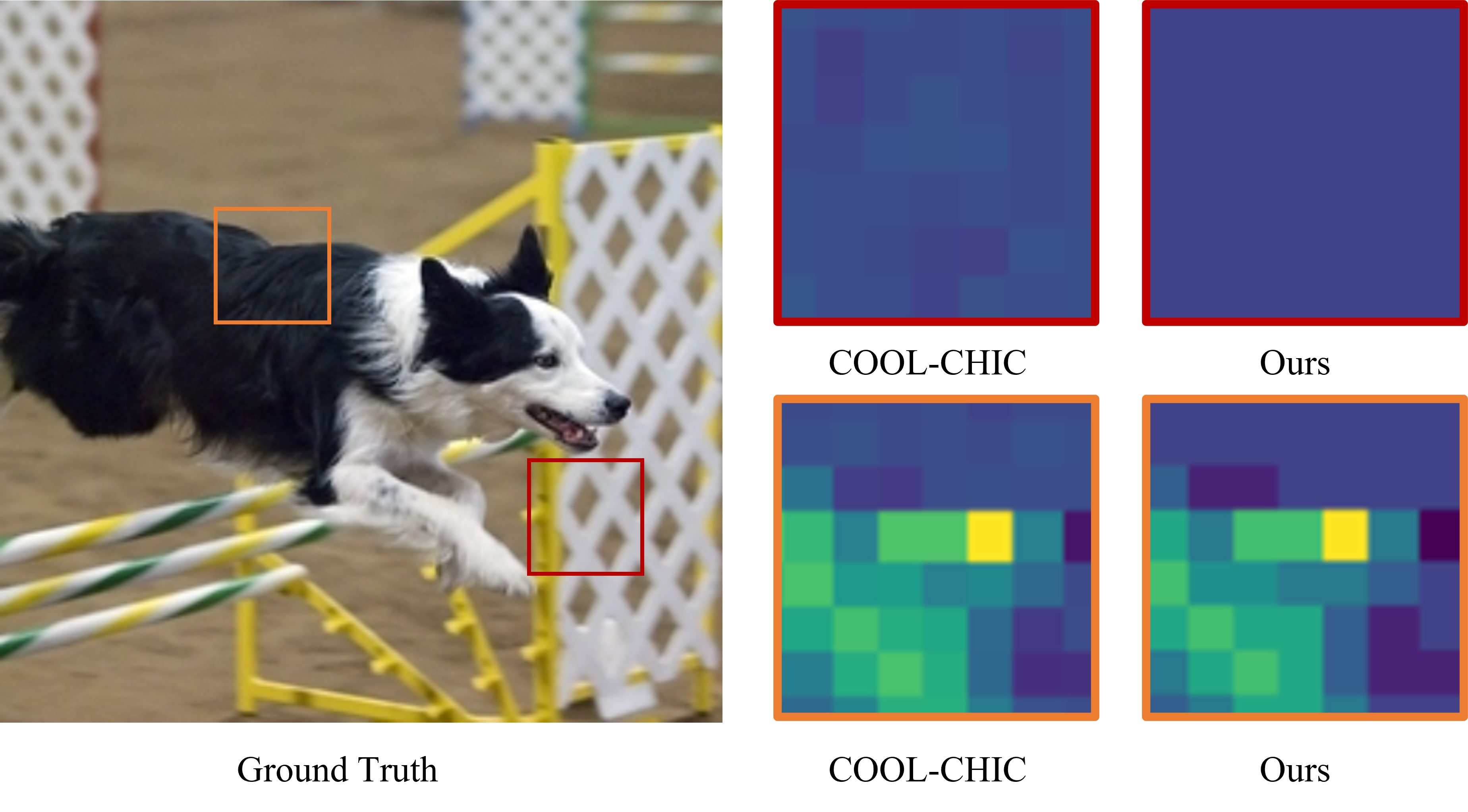}%
		\label{fig:visual1}
	}% 关键：删除换行，用%注释掉行尾空白符
	\subfloat{%
		\includegraphics[width=0.5\textwidth]{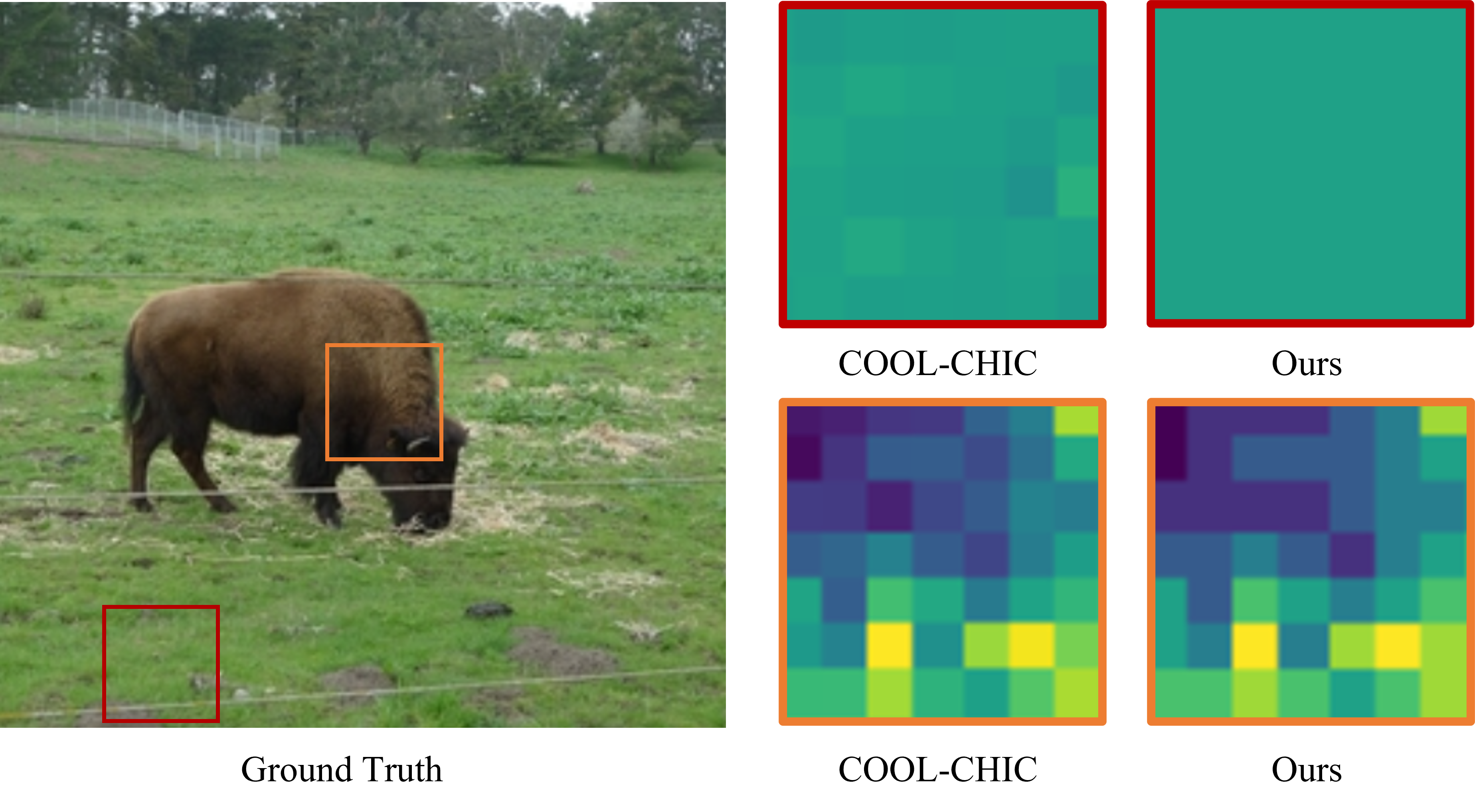}%
		\label{fig:visual2}
	}
	\caption{Visualization of features for image classification. For each example, the left image is the ground-truth image, where the red and orange boxes indicate two selected regions. The right images present the corresponding feature visualizations produced by COOL-CHIC and our method, respectively, where the background features are simplified and important features are preserved. }
	%Visualization of the features for the image classification,including (a) ground truth images, (b) features of COOL-CHIC, (c) features of IGDA, where the background features are simplified.}
\label{fig:visual_images}
\end{figure*}

\textbf{Evaluation on the IGFA.} We first analyze the influence of the architecture design in IGFA by comparing the LoRA‑based IGFA module (IGFA‑L), and the joint Kronecker product and LoRA‑based IGFA module (IGFA‑K). The results are shown in Table \ref{tab:igfaperformance}. It can be seen that the proposed IGFA-K significantly outperforms the IGFA-L. This is because the feature resolution is generally very large and only using the low-rank decomposition is not enough to reduce the number of parameters. Our IGFA-K can better explore the local properties of the features with the Kronecker product, and further reduce the number of parameters with low-rank approximation of the components used in the Kronecker product. 

Then similar comparison settings as in the above IGNA evaluation is used and the results are also shown in Table \ref{tab:igfaperformance}.
By comparing with the performance of the baseline and the single layer (SL) variants of IGFA-K, similar conclusions can be reached where our IGFA-K performs better than the baseline and SL variants when the number of parameters is large. Note that SL is only optimal under extremely low budgets, with only enough budgets to adapt one layer. In such a case, our IGFA or IGNA reduces to only optimizing one layer as illustrated in Section \ref{sec:igna}. When the budget is larger, our method performs much better. {It is worth noting that the adaptation parameters are quantized and entropy-coded in the INR codec. As a result, the actual bitrate overhead is lower than that implied by the raw parameter count. }
%修改修改

\begin{table}[!t]
	\centering
	\caption{Results of the IGDA, in comparison with only using IGFA, IGNA, and simply combining them as IGFA+IGNA.}
	\label{tab:igdaperformance}
	\renewcommand{\arraystretch}{1.2}
	\begin{tabular*}{\columnwidth}{@{\extracolsep{\fill}}>{\centering\arraybackslash}p{2cm}|ccc}
		\hline
		\multicolumn{1}{c|}{Method} & {Trainable param} & Bpp & Accuracy (\%) \\
		\hline
		baseline & - & 0.375 & 77.29 \\
		\hline
		IGNA & 1020 & 0.406 & 96.64 \\
		IGFA & 1168 & 0.410 & 96.90 \\
		IGFA+IGNA & 1151 & 0.410 & 96.64 \\
		IGDA & 973 & 0.402 & 97.80 \\
		\hline
	\end{tabular*}
\end{table}

\textbf{Evaluation on the IGDA with both IGFA and IGNA.}  The IGFA and IGNA can also be used together to form a dual adaptation. The results of using IGFA, IGNA, simply combining IGFA and IGNA with a similar number of parameters (noted as IGFA+IGNA), and IGDA are shown in Table \ref{tab:igdaperformance}.  All the models are using a similar number of parameters, and experiments are conducted under the Lagrangian multiplier 0.010 to provide a high quality.
It can be observed that the proposed IGDA achieves the best performance, demonstrating the benefit of importance‑guided adaptive design. Especially comparing with IGFA+IGNA, it can be seen that the importance guided allocation in our IGDA further leads to a performance improvement, with IGDA achieving 97.80\% accuracy compared to 96.64\% for IGFA+IGNA, validating the efficacy of our dual adaptation.
%the performance is significantly better since simply combining IGFA and IGNA cannot further increase the performance. 
Moreover, the performance improvement of IGDA is relatively small when compared with the IGFA and IGNA, mostly because the model under this specific Lagrangian multiplier reaches its limit due to quantization. Also it indicates that IGFA and IGNA are complementary to each other to some extent, and under a small bit budget, each one can be used to perform the adaptation task.

\textbf{Visual analysis of the learned features.} The visualization results of the feature produced by the layer with the highest importance score are shown in Fig. \ref{fig:visual_images}. It can be observed that the feature for human vision preserves the background details such as the grass, while our method effectively focuses on the semantic regions corresponding to the target classification object and suppresses irrelevant background information. This validates that the proposed method can adapt the model to extract relevant features useful for the machine vision task.

\section{Conclusion}

This paper proposes an Importance-Guided Dual feature and network Adaptation (IGDA) framework for ICMH. It explicitly explores how different features and network layers contribute to different  vision task performance, providing a principled foundation for efficient adaptation. Two adaptation modules, including the Importance Guided Network Adaptation and the Importance Guided Feature Adaptation are developed for network layer and feature adaptations, respectively. They can also be jointly optimized under a constrained parameter budget based on their different effects to the final performance. Extensive experiments on representative machine vision tasks demonstrate that the proposed IGDA framework achieves better rate–accuracy performance compared with existing ICMH methods. 

%\nocite{ladune2026cool}
\bibliographystyle{IEEEtran}
\bibliography{references}
 
  \begin{IEEEbiography}[{\includegraphics[width=1in,clip,keepaspectratio]{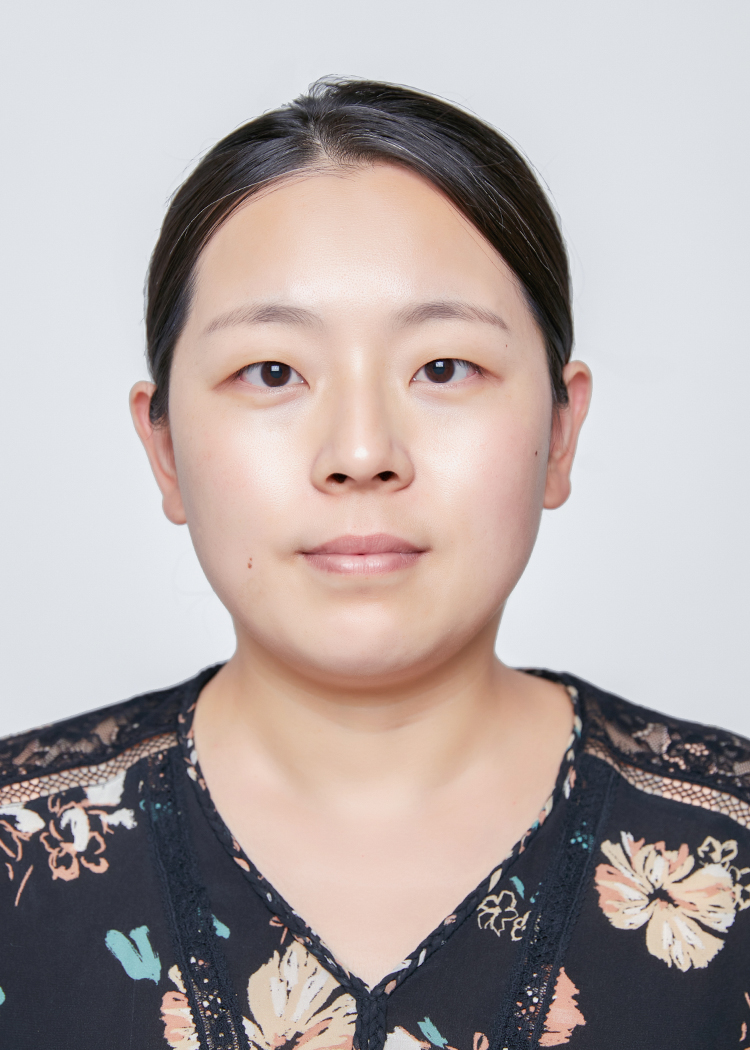}}]{Yanbo Gao}
 	is currently with the School of Software, Shandong University (SDU), Jinan, China, as an Associate Professor. She was with the School of Information and Communication Engineering, University of Electronic Science and Technology of China (UESTC), Chengdu, China, as a Post-doctor from 2018-2020. She received her Ph.D. degree from UESTC in 2018. Her research interests include video coding and 3D video processing. She was a co-recipient of the best student paper awards at the IEEE BMSB 2018.
 \end{IEEEbiography}
 
 \begin{IEEEbiography}[{\includegraphics[width=1in,clip,keepaspectratio]{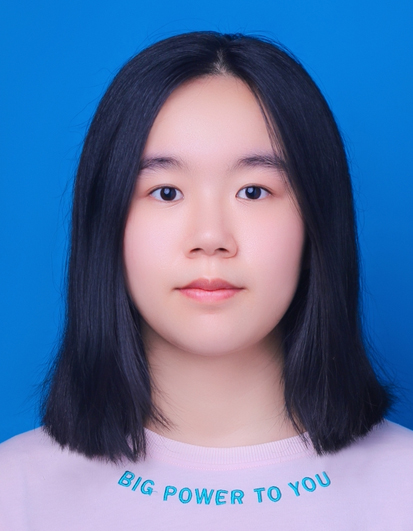}}]{Xinyu Zhao}
 	is currently pursuing the M.S. degree in Software Engineering in Shandong University. She received the B.S. degree in Software Engineering from Shandong University, Jinan, China, in 2024. Her research interests include deep learning and image compression.
 \end{IEEEbiography}
 \vspace{-0.8cm}
 \begin{IEEEbiography}[{\includegraphics[width=1in,clip,keepaspectratio]{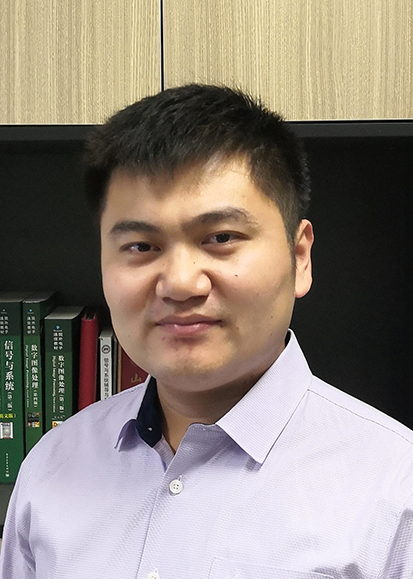}}]{Shuai Li}
 	is currently with the School of Control Science and Engineering, ShanDong University (SDU), China, as a Professor and QiLu Young Scholar. He was with the School of Information and Communication Engineering, University of Electronic Science and Technology of China, China, as an Associate Professor from 2018-2020. He received his Ph.D. degree from the University of Wollongong, Australia, in 2018. His research interests include image/video coding, 3D video processing and computer vision. He has received three Best Paper Awards at the ISCAS 2025, IEEE BMSB 2014 and IIH-MSP 2013, respectively.
 \end{IEEEbiography}
  \vspace{-0.8cm}
 \begin{IEEEbiography}[{\includegraphics[width=0.95in,clip,keepaspectratio]{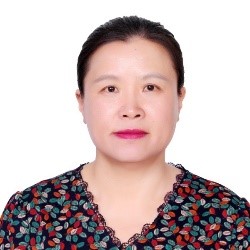}}]{Xun Cai}
 	is currently with the School of Software, Shandong University (SDU), China, as an Associate Professor. She received her Ph.D. degree from Shandong University in 2007. She has been to the University of Texas at Arlington in the US as a nationally assigned scholar-visitor from 2009 to 2010. Her research interests include image/video recognition, computer vision, and industrial simulation.
 \end{IEEEbiography}
  \vspace{-0.8cm}
 \begin{IEEEbiography}[{\includegraphics[width=1in,clip,keepaspectratio]{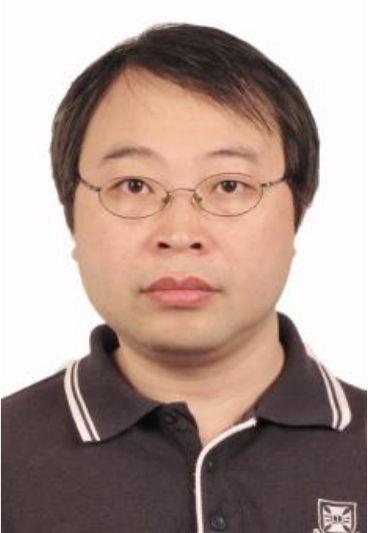}}]{Mao Ye}	 (Senior Member, IEEE) received the B.S. degree from Sichuan Normal University, Chengdu, China, in 1995, the M.S. degree from the University of Electronic Science and Technology of China, Chengdu, in 1998, and the Ph.D. degree from The Chinese University of Hong Kong, Hong Kong, in 2002, all in mathematics. He has been a short-time Visiting Scholar at the University of Queensland and the University of Pennsylvania. He is currently a Professor and the Director of the CVLab, University of Electronic Science and Technology of China, Chengdu. His research interests include machine learning and computer vision. In these areas, he has published over 90 articles in leading international journals or conference proceedings. He was a co-recipient of the Best Student Paper Award at the IEEE ICME 2017. He has served on the Editorial Board of the Engineering Applications of Artificial Intelligence.
 \end{IEEEbiography}
  \vspace{-0.8cm}
 \begin{IEEEbiography}[{\includegraphics[width=1in,height=1.25in,clip,keepaspectratio]{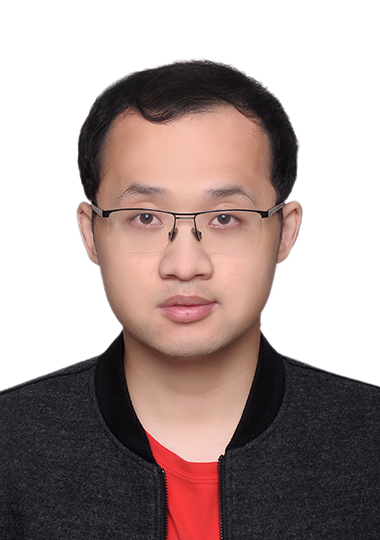}}] {Li Li} (M'17-SM'25) received the B.S. and Ph.D. degrees in electronic engineering from University of Science and Technology of China (USTC), Hefei, Anhui, China, in 2011 and 2016, respectively.
 	He was a visiting assistant professor in University of Missouri-Kansas City from 2016 to 2020.
 	He joined the department of electronic engineering and information science of USTC as a research fellow in 2020 and became a professor in 2022. \par
 	His research interests include image/video/point cloud coding and processing.
 	He has authored or co-authored more than 100 papers in international journals and conferences. 
 	He has more than 20 granted patents. 
 	He has several technique proposals adopted by standardization groups.
 	He received the Multimedia Rising Star 2023.
 	He received the Best 10\% Paper Award at the 2016 IEEE Visual Communications and Image Processing (VCIP) and the 2019 IEEE International Conference on Image Processing (ICIP).
 	He serves as a senior associate editor for \textsc{IEEE Transactions on Circuits and Systems for Video Technology} since 2026 and an associate editor for \textsc{IEEE Transactions on Multimedia} since 2025. 
 \end{IEEEbiography}
\vfill

\end{document}